%% file: nageom.tex
\include{header}
\documentclass[letterpaper,12pt]{article}

\begin{document}

\include{defs}

\include{title}

\newpage

\setcounter{page}{1}

\include{body}

\newpage

\section*{Acknowledgements}
The work of D.V.N. was partially supported by DOE grant
DE-F-G03-95-ER-40917.

\bibliographystyle{hunsrt}
\bibliography{bibdata}

\include{append}

\end{document}

%% file: defs.tex
\voffset0truein       
\hoffset0truein        
\tolerance = 10000

\newcommand{\comment}[1]{}
\newcommand{\eqn}[1]{Eq. (\ref{#1})}
\newcommand{\eqs}[2]{Eqs. (\ref{#1},\ref{#2})}

\setlength{\unitlength}{1mm}

\oddsidemargin  30pt
\evensidemargin  10pt
\textheight  615pt
\textwidth  460pt
\parskip 0.3cm
\headheight  0pt
\headsep  0pt
\footskip  40pt

\renewcommand{\thefootnote}{\fnsymbol{footnote}}
\setcounter{footnote}{0}
\newlength{\pubnumber} \settowidth{\pubnumber}{CLNS~99/9999~}
\newcommand\pubblock[2]{\begin{flushright}\parbox{\pubnumber}
 {\begin{flushleft}#1\\ #2\\ \end{flushleft}}\end{flushright}}
\catcode`\@=11
\@addtoreset{equation}{section}
\def\theequation{\thesection.\arabic{equation}}
\def\section{\@startsection{section}{1}{\z@}{3.5ex plus 1ex minus .2ex}
 {2.3ex plus .2ex}{\large\bf}}
\def\subsection{\@startsection{subsection}{2}{\z@}{2.3ex plus .2ex}
 {2.3ex plus .2ex}{\bf}}
\newcommand\Appendix[1]{\def\thesection{Appendix \Alph{section}}
 \section{\label{#1}}\def\thesection{\Alph{section}}}


\def\gappeq{\mathrel{\rlap {\raise.5ex\hbox{$>$}}
{\lower.5ex\hbox{$\sim$}}}}

\def\lappeq{\mathrel{\rlap{\raise.5ex\hbox{$<$}}
{\lower.5ex\hbox{$\sim$}}}}

\def\Toprel#1\over#2{\mathrel{\mathop{#2}\limits^{#1}}}
\def\FF{\Toprel{\hbox{$\scriptscriptstyle(-)$}}\over{$\nu$}}

\def\beq{\begin{equation}}
\def\eeq{\end{equation}}
\def\bea{\begin{eqnarray}}
\def\eea{\end{eqnarray}}
\def\bq{\begin{quote}}
\def\eq{\end{quote}}
\def\ie{i.e.\ }
\def\eg{e.g.\ }
\def\cf{c.f.\ }
\def\tetal{et~al.\ }
\def\ala{{\it \`{a}~la}\ }

\def\vev#1{\langle #1\rangle}
\def\r#1{$\bf#1$}
\def\rb#1{$\bf\overline{#1}$}
\def\AJ{{\it Astrophys.J.} }
\def\AJL{{\it Ap.J.Lett.} }
\def\AJS{{\it Ap.J.Supp.} }
\def\AM{{\it Ann.Math.} }
\def\AP{{\it Ann.Phys.} }
\def\APJ{{\it Ap.J.} }
\def\APP{{\it Acta Phys.Pol.} }
\def\ASAS{{\it Astron. and Astrophys.} }
\def\BAMS{{\it Bull.Am.Math.Soc.} }
\def\CMJ{{\it Czech.Math.J.} }
\def\CMP{{\it Commun.Math.Phys.} }
\def\FP{{\it Fortschr.Physik} }
\def\HPA{{\it Helv.Phys.Acta} }
\def\IJMP{{\it Int.J.Mod.Phys.} }
\def\JMM{{\it J.Math.Mech.} }
\def\JP{{\it J.Phys.} }
\def\JCP{{\it J.Chem.Phys.} }
\def\LNC{{\it Lett. Nuovo Cimento} }
\def\SNC{{\it Suppl. Nuovo Cimento} }
\def\MPL{{\it Mod.Phys.Lett.} }
\def\NAT{{\it Nature} }
\def\NC{{\it Nuovo Cimento} }
\def\NP{{\it Nucl.Phys.} }
\def\PL{{\it Phys.Lett.} }
\def\PR{{\it Phys.Rev.} }
\def\PRL{{\it Phys.Rev.Lett.} }
\def\PRTS{{\it Physics Reports} }
\def\PS{{\it Physica Scripta} }
\def\PTP{{\it Progr.Theor.Phys.} }
\def\RMPA{{\it Rev.Math.Pure Appl.} }
\def\RNC{{\it Rivista del Nuovo Cimento} }
\def\SJPN{{\it Soviet J.Part.Nucl.} }
\def\SP{{\it Soviet.Phys.} }
\def\TMF{{\it Teor.Mat.Fiz.} }
\def\TMP{{\it Theor.Math.Phys.} }
\def\YF{{\it Yadernaya Fizika} }
\def\ZETF{{\it Zh.Eksp.Teor.Fiz.} }
\def\ZP{{\it Z.Phys.} }
\def\ZMP{{\it Z.Math.Phys.} }
\def\AEF{A.E. Faraggi}
\def\AP#1#2#3{{\it Ann.\ Phys.}\/ {\bf#1} (#2) #3}
\def\EPJC#1#2#3{{\it The Euro.\ Phys.\ Jour.\/} {\bf C#1} (#2) #3}
\def\NPB#1#2#3{{\it Nucl.\ Phys.}\/ {\bf B#1} (#2) #3}
\def\NPBPS#1#2#3{{\it Nucl.\ Phys.}\/ {{\bf B} (Proc. Suppl.) {\bf #1}} (#2) 
 #3}
\def\PLB#1#2#3{{\it Phys.\ Lett.}\/ {\bf B#1} (#2) #3}
\def\PRD#1#2#3{{\it Phys.\ Rev.}\/ {\bf D#1} (#2) #3}
\def\PRL#1#2#3{{\it Phys.\ Rev.\ Lett.}\/ {\bf #1} (#2) #3}
\def\PRT#1#2#3{{\it Phys.\ Rep.}\/ {\bf#1} (#2) #3}
\def\PTP#1#2#3{{\it Prog.\ Theo.\ Phys.}\/ {\bf#1} (#2) #3}
\def\MODA#1#2#3{{\it Mod.\ Phys.\ Lett.}\/ {\bf A#1} (#2) #3}
\def\MPLA#1#2#3{{\it Mod.\ Phys.\ Lett.}\/ {\bf A#1} (#2) #3}
\def\IJMP#1#2#3{{\it Int.\ J.\ Mod.\ Phys.}\/ {\bf A#1} (#2) #3}
\def\IJMPA#1#2#3{{\it Int.\ J.\ Mod.\ Phys.}\/ {\bf A#1} (#2) #3}
\def\nuvc#1#2#3{{\it Nuovo Cimento}\/ {\bf #1A} (#2) #3}
\def\RPP#1#2#3{{\it Rept.\ Prog.\ Phys.}\/ {\bf #1} (#2) #3}
\def\etal{{\it et al\/}}

\def\tfrac#1#2{{\textstyle \frac{#1}{#2}}}
\def\IC{\relax\hbox{$\inbar\kern-.3em{\rm C}$}}
\def\IQ{\relax\hbox{$\inbar\kern-.3em{\rm Q}$}}
\def\IR{\relax{\rm I\kern-.18em R}}
 \font\cmss=cmss10 \font\cmsss=cmss10 at 7pt
 \font\cmsst=cmss10 at 9pt
 \font\cmssn=cmss9

\def\IZ{\relax\ifmmode\mathchoice
 {\hbox{\cmss Z\kern-.4em Z}}{\hbox{\cmss Z\kern-.4em Z}}
 {\lower.9pt\hbox{\cmsss Z\kern-.4em Z}}
 {\lower1.2pt\hbox{\cmsss Z\kern-.4em Z}}\else{\cmss Z\kern-.4em Z}\fi}

\def\Io{\relax\ifmmode\mathchoice
 {\hbox{\cmss 1\kern-.4em 1}}{\hbox{\cmss 1\kern-.4em 1}}
 {\lower.9pt\hbox{\cmsss 1\kern-.4em 1}}
 {\lower1.2pt\hbox{\cmsss 1\kern-.4em 1}}\else{\cmss 1\kern-.4em 1}\fi}

\def\ZX{{Z \hskip -8pt Z}}

\def\ZZ{\relax\ifmmode\mathchoice
 {\hbox{\cmss Z\kern-.4em Z}}{\hbox{\cmss Z\kern-.4em Z}}
 {\lower.9pt\hbox{\cmsss Z\kern-.4em Z}}
 {\lower1.2pt\hbox{\cmsss Z\kern-.4em Z}}\else{\cmss Z\kern-.4em Z}\fi}

\def\11{\relax\ifmmode\mathchoice
 {\hbox{\cmss 1\kern-.4em 1}}{\hbox{\cmss 1\kern-.4em 1}}
 {\lower.9pt\hbox{\cmsss 1\kern-.4em 1}}
 {\lower1.2pt\hbox{\cmsss 1\kern-.4em 1}}\else{\cmss 1\kern-.4em 1}\fi}

\def\lam#1{\lambda_{#1}}
\def\non{\nonumber}
\def\smgg{ $SU(3)_C\times SU(2)_L\times U(1)_Y$ }
\def\smggb{ $SU(3)_C\times SU(2)_L\times U(1)_Y$}
\def\SM{Standard-Model }
\def\SUSY{supersymmetry }
\def\SSSM{supersymmetric standard model}
\def\MSSM{minimal supersymmetric standard model}
\def\MSSSM{MS$_{str}$SM }
\def\MSSSMc{MS$_{str}$SM, }
\def\obs{{\rm observable}}
\def\sig{{\rm singlets}}
\def\hid{{\rm hidden}}
\def\MS{M_{str}}
\def\Ms{$M_{str}$}
\def\MP{M_{P}}
\def\GeV{\,{\rm GeV}}
\def\TeV{\,{\rm TeV}}

\def\at{ }
\def\vev#1{\langle #1\rangle}
\def\mvev#1{|\langle #1\rangle|^2}
\def\mveV#1{|\langle #1\rangle|}
\def\ket#1{|#1\rangle}

\def\UA{U(1)_{\rm A}}
\def\QA{Q^{(\rm A)}}
\def\mssm{SU(3)_C\times SU(2)_L\times U(1)_Y} 

\def\NA{non-Abelian }
\def\KM{Ka\v c-Moody }

\def\eps{\epsilon}
\def\fhalf{\frac{1}{2}}
\def\fsqrt{\frac{1}{\sqrt{2}}}
\def\half{{\textstyle{1\over 2}}}
\def\third{{\textstyle {1\over3}}}
\def\quarter{{\textstyle {1\over4}}}
\def\sixth{{\textstyle {1\over6}}}

\def\op#1{$\Phi_{#1}$}
\def\opp#1{$\Phi^{'}_{#1}$}
\def\opb#1{$\overline{\Phi}_{#1}$}
\def\opbp#1{$\overline{\Phi}^{'}_{#1}$}
\def\oppb#1{$\overline{\Phi}^{'}_{#1}$}
\def\oppx#1{$\Phi^{(')}_{#1}$}
\def\opbpx#1{$\overline{\Phi}^{(')}_{#1}$}

\def\oh#1{$h_{#1}$}
\def\ohb#1{${\bar{h}}_{#1}$}
\def\ohp#1{$h^{'}_{#1}$}

\def\oQ#1{$Q_{#1}$}
\def\odc#1{$d^{c}_{#1}$}
\def\ouc#1{$u^{c}_{#1}$}

\def\oL#1{$L_{#1}$}
\def\oec#1{$e^{c}_{#1}$}
\def\oNc#1{$N^{c}_{#1}$}

\def\oH#1{$H_{#1}$}
\def\oV#1{$V_{#1}$}
\def\oHs#1{$H^{s}_{#1}$}
\def\oVs#1{$V^{s}_{#1}$}

\def\p4{\Phi_4}
\def\pp4{\Phi^{'}_4}
\def\pb4{\bar{\Phi}_4}
\def\ppb4{\bar{\Phi}^{'}_4}
\def\p#1{{\Phi_{#1}}}
\def\P#1{{\Phi_{#1}}}
\def\pp#1{{\Phi^{'}_{#1}}}
\def\pb#1{{{\overline{\Phi}}_{#1}}}
\def\bp#1{{{\overline{\Phi}}_{#1}}}
\def\pbp#1{{{\overline{\Phi}}^{'}_{#1}}}
\def\ppb#1{{{\overline{\Phi}}^{'}_{#1}}}
\def\bpp#1{{{\overline{\Phi}}^{'}_{#1}}}
\def\bi#1{{{\overline{\Phi}}^{'}_{#1}}}
\def\ppx#1{{\Phi^{(')}_{#1}}}
\def\pbpx#1{{\overline{\Phi}^{(')}_{#1}}}

\def\h#1{h_{#1}}
\def\hb#1{{\bar{h}}_{#1}}
\def\hp#1{h^{'}_{#1}}

\def\Q#1{Q_{#1}}
\def\dc#1{d^{c}_{#1}}
\def\uc#1{u^{c}_{#1}}

\def\L#1{L_{#1}}
\def\ec#1{e^{c}_{#1}}
\def\Nc#1{N^{c}_{#1}}

\def\H#1{H_{#1}}
\def\V#1{V_{#1}}
\def\Hs#1{{H^{s}_{#1}}}
\def\sH#1{{H^{s}_{#1}}}
\def\Vs#1{{V^{s}_{#1}}}
\def\sV#1{{V^{s}_{#1}}}

\def\fdtv{FD2V }
\def\fdtp{FD2$^{'}$ }
\def\fdtpv{FD2$^{'}$v }

\def\FD2pv{FD2$^{'}$V }
\def\FD2p{FD2$^{'}$ }

\def\zz{$Z_2\times Z_2$ }

\def\Tr{{\rm Tr}\, }
\def\tr{{\rm tr}\, }

\def\y{\,{\rm y}}
\def\l{\langle}
\def\r{\rangle}
\def\o#1{\frac{1}{#1}}

\def\zi{z_{\infty}}

\def\bc{\bar{c}}
\def\lh{\bar{h}}
\def\hb#1{\bar{h}_{#1}}
\def\bh#1{\bar{h}_{#1}}
\def\Htw{{\tilde H}}
\def\chibar{{\overline{\chi}}}
\def\qbar{{\overline{q}}}
\def\ibar{{\overline{\imath}}}
\def\jbar{{\overline{\jmath}}}
\def\Hbar{{\overline{H}}}
\def\Qbar{{\overline{Q}}}
\def\abar{{\overline{a}}}
\def\alphabar{{\overline{\alpha}}}
\def\betabar{{\overline{\beta}}}
\def\tautwo{{ \tau_2 }}
\def\thetatwo{{ \vartheta_2 }}
\def\thetathree{{ \vartheta_3 }}
\def\thetafour{{ \vartheta_4 }}
\def\ttwo{{\vartheta_2}}
\def\tthree{{\vartheta_3}}
\def\tfour{{\vartheta_4}}
\def\ti{{\vartheta_i}}
\def\tj{{\vartheta_j}}
\def\tk{{\vartheta_k}}
\def\calF{{\cal F}}
\def\smallmatrix#1#2#3#4{{ {{#1}~{#2}\choose{#3}~{#4}} }}
\def\ab{{\alpha\beta}}
\def\Minv{{ (M^{-1}_\ab)_{ij} }}
\def\ii{{(i)}}
\def\V{{\bf V}}
\def\N{{\bf N}}

\def\b{{\bf b}}
\def\S{{\bf S}}
\def\X{{\bf X}}
\def\I{{\bf I}}
\def\bone{{\mathbf 1}}
\def\bo{{\mathbf 0}}
\def\bs{{\mathbf S}}
\def\mS{{\mathbf S}}
\def\bS{{\mathbf S}}
\def\bb{{\mathbf b}}
\def\mb{{\mathbf b}}
\def\mX{{\mathbf X}}
\def\mI{{\mathbf I}}
\def\bI{{\mathbf I}}
\def\balpha{{\mathbf \alpha}}
\def\bbeta{{\mathbf \beta}}
\def\bgamma{{\mathbf \gamma}}
\def\bxi{{\mathbf \xi}}
\def\malpha{{\mathbf \alpha}}
\def\mbeta{{\mathbf \beta}}
\def\mgamma{{\mathbf \gamma}}
\def\mxi{{\mathbf \xi}}
\def\bphi{\overline{\Phi}}

\def\eps{\epsilon}

\def\t#1#2{{ \Theta\left\lbrack \matrix{ {#1}\cr {#2}\cr }\right\rbrack }}
\def\C#1#2{{ C\left\lbrack \matrix{ {#1}\cr {#2}\cr }\right\rbrack }}
\def\tp#1#2{{ \Theta'\left\lbrack \matrix{ {#1}\cr {#2}\cr }\right\rbrack }}
\def\tpp#1#2{{ \Theta''\left\lbrack \matrix{ {#1}\cr {#2}\cr }\right\rbrack }}
\def\l{\langle}
\def\r{\rangle}

%% file: title.tex
\begin{titlepage}
\setcounter{page}{1}
\rightline{MIFP-05-32, ACT-05-12, BU-HEPP-05-10, CASPER-05-11}
\rightline{\tt hep-th/0512020}
\rightline{December 2005}
\vspace{.06in}
\begin{center}
{\Large \bf On Geometrical Interpretation\\
	of Non-Abelian
	Flat Direction Constraints}

\vspace{.25in}
{\large
        G.B. Cleaver,$^{2,3}$\footnote{gerald{\underline{\phantom{a}}}cleaver@baylor.edu}
        D.V. Nanopoulos,$^{1,2,4}$\footnote{dimitri@physics.tamu.edu}\\
        J.T. Perkins,$^{3}$\footnote{john{\underline{\phantom{a}}}perkins@baylor.edu}
        and J.W. Walker$^{5}$\footnote{jwalker@shsu.edu}}
\\
\vspace{1in}
{\it $^{1}$ George~P. and Cynthia~W. Mitchell Institute of Fundamental Physics,\\
	    Texas A\&M University,\\
	    College Station, TX 77843-4242, USA\\}
\vspace{.06in}
{\it $^{2}$ Astro Particle Physics Group,\\
            Houston Advanced Research Center (HARC),
            The Mitchell Campus,
            Woodlands, TX 77381, USA\\}
\vspace{.06in}
{\it$^{3}$  Center for Astrophysics, Space Physics \& Engineering Research\\
            Department of Physics, Baylor University,
            Waco, TX 76798-7316, USA\\}
\vspace{.06in}
{\it$^{4}$  Academy of Athens, Chair of Theoretical Physics,\\ 
            Division of Natural Sciences,\\
            28 Panepistimiou Avenue, Athens 10679, Greece\\}
\vspace{.06in}
{\it$^{5}$  Sam Houston State University,\\
            Visiting Assistant Professor, Department of Physics,\\
            Huntsville, TX 77341, USA\\}
\end{center}

\newpage
\begin{abstract}
In order to produce a low energy effective field theory from a string model, it is necessary to specify a vacuum state.
In order that this vacuum be supersymmetric, it is well known that all field expectation values must be along so-called 
flat directions, leaving the $F$- and $D$-terms of the scalar potential to be zero. The situation becomes particularly 
interesting  when one attempts to realize such directions while assigning VEVS to fields transforming under non-Abelian 
representations of the gauge group. Since the expectation value is now shared among multiple components of a field, 
satisfaction of flatness becomes an inherently geometrical problem in the group space. Furthermore, the possibility 
emerges that a single seemingly dangerous $F$-term might experience a self-cancellation among its components.
The hope exists that the geometric language can provide an intuitive and immediate recognition of when the $D$ and $F$ 
conditions are simultaneously compatible, as well as a powerful tool for their comprehensive classification.
This is the avenue explored in this paper, and applied to the cases of $SU(2)$ and $SO(2N)$, relevant respectively 
to previous attempts at reproducing the MSSM and the flipped $SU(5)$ GUT.
Geometrical interpretation of non-Abelian flat directions finds application to M-theory through the recent 
conjecture of equivalence between $D$-term strings and wrapped $D$-branes of Type II theory\cite{Dvali:2003zh}. 
Knowledge of the geometry of the flat direction ``landscape'' of a $D$-term string model could yield information 
about the dual brane model.
It is hoped that the techniques encountered will be of benefit in extending the viability of the quasi-realistic 
phenomenologies already developed.
\end{abstract}

\smallskip
\end{titlepage}

%% file: body.tex
\section{Introduction}

\subsection{The Task at Hand}

The central task of physics has always been the distillation of information down into knowledge;
the union of facts seemingly disparate, by common woven threads of principle.
The revelation of modern physics has been that the proper eye with which to perceive this tapestry
is that of the ultra-high energy micro-world.
The convergence of gauge couplings speaks again to complexity rising out of simplicity at a scale
alluringly close to the expected regime of quantum gravity.
The satisfaction of low-energy constraints \cite{Ellis:1990zq,Amaldi:1991cn,Langacker:1991an}
while using only MSSM field content\footnote{Standard Model here also refers to its extensions
that include neutrino masses.} reveals the vast desert for what it is \cite{Ghilencea:2001qq}, a great
and empty field for the couplings to run in and produce in their crossing our familiar world.
However, spanning such unimaginable leagues requires a fundamental shift of paradigm, and tools
born to the world in ways unfamiliar.
String theory and its relatives, M-theory and local supersymmetry, present mathematics as the
new microscope, and point to our new guides for the unexplored realm: beauty,
simplicity and above all else, unity.
Great promise exists in the natural, consistent and necessary accommodation of gravity,
as well as the tremendous gauge freedom and large symmetry groups which may be transferred out of
the intrinsically resident extra spatial dimensions.
Great shortcoming exists in the same places, as the theory seems in fact overly malleable,
and incapable of definite predictions.
Surely predictivity is not everything though, and especially in this new climate, insight,
context and origin for knowledge previously held is a great achievement in itself.
By the same token the door is held open here for a type of phenomenology whereby known effects,
chiefly supersymmetry, are used to limit the possible parameter space.
While non-perturbative effects must finally produce a dynamic and definite solution \cite{Witten:1996mz},
it seems there is also a value to searching in the available light of a preferred theoretical
framework and constraint for the simple existence of a viable, albeit ad-hoc, solution.
It is hoped the techniques so developed will have enduring relevance, and that the examples
unearthed may illuminate some properties of the true path, serving to narrow the gulf from
the low-energy side, and set a table in waiting for the weary travelers who cross the desert.

\subsection{Heterotic String Model Building}

In the context of Heterotic string theory there has historically been reasonably good
success in the construction of both $MSSM$ and Flipped $SU(5)$ models by the fixing of
phases on the fermionic degrees of freedom.
In order to produce a low energy effective field theory from a string model,
it is necessary to further specify a vacuum state.
The Fayet-Illiopoulos (FI) anomaly which was originally proposed as a mechanism for
electroweak SUSY breaking
\footnote{This approach has generally been supplanted by the writing of `soft'
supersymmetry breaking terms in the Lagrangian which are supposed in turn to
descend dynamically from spontaneously broken no-scale supergravity.}
has been recast here to a dignified position at the string scale.
Firstly, it helps greatly to remove the extraneous $U(1)$ factors which tend to appear
abundantly in such constructions, and secondly demands the assignment of a non-trivial
set of vacuum expectation values (VEVs) which will cancel the offending term such
that string-scale SUSY now be preserved!

The door is held open here for a type of phenomenology whereby preservation of
supersymmetry is used as the foremost constraint to limit the possible parameter space.
In order that the vacuum respect SUSY, all field
expectation values must be along so-called `flat directions', leaving the
$F$- and $D$-terms of the scalar potential to be zero.
It is {\it from} these required assignments that a beneficial spectral truncation
and values for masses and couplings may be realized.
We note also in passing that world sheet selection rules form a new criteria for
elimination of superpotential terms above simple gauge invariance.

The situation becomes particularly interesting when one attempts to realize
flat directions while assigning VEVS to fields transforming under non-Abelian
representations of the gauge group.
Such a process has been suggested by the insufficiency to date of simpler
Abelian-only constructions to generate sufficient mass terms,
among other shortcomings. 
Since the expectation value is now shared among multiple components of a
field, satisfaction of flatness becomes an inherently geometrical problem
in the group space.
Specifically it has been noticed that the $D$-term $SUSY$ condition appears becomes
translated into the imperative that the adjoint space representation of all expectation
values form a closed vector sum.
Furthermore, the possibility emerges that a single seemingly dangerous
$F$-term might experience a self-cancellation among its components.
The potential exists that this geometric language can provide an intuitive and
immediate recognition of when the $D$ and $F$ conditions are simultaneously
compatible, as well as a powerful tool for their comprehensive classification.
This is the avenue receiving the greatest attention in the present section,
as applied to the cases of $SU(2)$ and $SO(2n)$.
These are relevant respectively to previous attempts at reproducing the FNY
MSSM and the flipped $SU(5)$ GUT by way of its confining hidden sector element
$SU(4) \sim SO(6)$.
By necessity, the second case addresses the issues inherent to groups with
rank $n>1$.
An additional elimination of some otherwise dangerous terms from the superpotential
already been achieved in this formalism.
It is hoped that the techniques encountered will be of further benefit in extending
the viability of the quasi-realistic phenomenologies already developed.

\subsection{Minimal Superstring Standard Models (MSSM's)} 

The approach we favor here for model production uses free fermions 
for the internal degrees of freedom of a 
heterotic string theory,\cite{Antoniadis:1986rn,Kawai:1986ah} as in the ``FNY'' model of
ref.\ \cite{Faraggi:1989ka,Faraggi:1990af} 
with its $SU(3)_C \times SU(2)_L \times U(1)_Y \times \prod_i U(1)_i$ observable gauge group.
Preservation of supersymmetry mandates the acquisition of non-zero vacuum expectation values (VEVs) to cancel the Fayet-Iliopoulos (FI) $D$-term, which arises in conjunction with the anomalous $U(1)$ factor endemic to many related constructions, while keeping all other $D$- and $F$-terms zero as well.
In this context, the assignment of a satisfactory ground state is a delicate and confining business, but some freedom does remain to tailor phenomenology in the emerging low energy effective field theory.
Success has been had here with VEVs decoupling all Standard Model charged fields outside the MSSM \cite{Cleaver:1999xk,Cleaver:1999cj,Cleaver:1999mw,Cleaver:2000aa}, and mechanisms of generational mass suppression have arisen from powers of $\frac{\vev{\phi}}{\MP}$ with non-renormalizable terms and from coupling to Higgs fields with differing contributions to the massless physical combination.
Promising, if imperfect, particle properties have been realized and discussed, but indications exist from varied directions \cite{Cleaver:2000aa,Cleaver:2001ab,Lopez:1991ac,Antoniadis:1991fc,Faraggi:1992yz,Faraggi:1993sg,Faraggi:1993su} that attempts restricted to a \NA singlet vacuum are unsatisfactory, and that perhaps nature's craft avails a larger set of tools.
This paper will then focus generally on the technology of assigning VEVs to \NA fields, and in particular on the geometrical framework that is introduced by the presence of VEV components within a group space.
The geometrical point of view facilitates manipulations which emerge for \NA VEVs, such as treating superpotential contractions with multiple pairings, and examining the new possibility of self-cancellation between elements of a single term.
When expressed in this language, the process of describing valid solutions, or compatibilities between the $F$ and $D$ conditions, can become more accessible and intuitive, and will hopefully aid in closing the gap between string model building and low energy experimental evidence.

The sequence taken will be first to review the constraints which supersymmetry imposes on our VEV choices, in particular 
for the \NA case.
Next, the discussion will be made concrete by turning to the case of $SU(2)$, and following that, $SO(2n)$.
These choices are made because of their application to the string-derived FNY 
MSSM, and flipped $SU(5)$ GUT \cite{Antoniadis:1989zy,Lopez:1992kg}, via $SO(6)$, respectively, but other benefits exist 
as well. $SU(2)$ is well known from the theory of spin-$\frac{1}{2}$ systems, and as a rank $1$ group with a number of 
generators equal to its fundamental dimension ($3$), it represents the simplest specific case on which to initiate 
discussion. $SO(2n)$, on the other hand, will generally represent one of the four main Lie group classifications, and 
introduces new complications by way of higher rank groups and an adjoint space of dimension greater than the fundamental.
Furthermore, although the fields under consideration are all space-time scalars, superpotential terms can inherit an 
induced symmetry property from the analytic rotationally invariant contraction form of the group under study.
$SU(2)$ will have a ``fermionic'' nature, with an antisymmetric contraction, while that of $SO(2n)$ will be symmetric, 
or ``bosonic''. 
 
Development of systematic methods for geometrical analysis of the landscape of $D$- and $F$-flat 
directions, as begun here, offers a possibility for further understanding of the geometry of brane-anti-brane systems.
This is related to the conjectured connection between 4D supergravity $D$-terms in the low energy effective 
field theory of a string and a $D_{3+q}-\bar{D}_{3+q}$ wrapped brane-anti-brane system. In this association, 
the energy of a $D_{3+q}-\bar{D}_{3+q}$ system appears as an FI $D$-term, an open string tachyon connecting
brane and anti-brane is revealed as an FI-cancelling Higgs field, and a $D_{1+q}$-brane produced in an  
annihilation between a $D_{3+q}$-brane and a $\bar{D}_{3+q}$-anti-brane is construed to be a $D$-term string 
\cite{Dvali:2003zh}. 

Interspersed in the document body will be special topics, such as self-cancellation, and 
specific examples of 
superpotential terms. The final section will be concluding remarks. The generators of $SO(6)$ are provided
in the appendix.

\section{$D$- and $F$-Flatness Constraints} 

The well known requirements for preservation of space-time supersymmetry, as expressed in the so-called $F$ and $D$-terms 
have been reviewed in \cite{Cleaver:1999xk,Cleaver:1999cj,Cleaver:1999mw,Cleaver:2000aa,Cleaver:2001ab}.
They will again be summarized here,\footnote{Portions of this section, most 
notably regarding flatness constraints and the group $SU(2)$, have been previously
reported in ref.\ \cite{Cleaver:2001ab}.},
with a new emphasis on geometric interpretation of the non-Abelian VEVs.

Space-time supersymmetry is broken in a model when the expectation value of the scalar potential,
\bea
 V(\varphi) = \half \sum_{\alpha} g_{\alpha}^2 
(\sum_{a=1}^{{\rm {dim}}\, ({\cal G}_{\alpha})} D_a^{\alpha} D_a^{\alpha}) +
                    \sum_i | F_{\varphi_i} |^2\,\, ,
\label{vdef}
\eea
becomes non-zero. 
The $D$-term contributions in (\ref{vdef}) have the form,   
\bea
D_a^{\alpha}&\equiv& \sum_m \varphi_{m}^{\dagger} T^{\alpha}_a \varphi_m\,\, , 
\label{dtgen} 
\eea
with $T^{\alpha}_a$ a matrix generator of the gauge group ${\cal G}_{\alpha}$ 
for the representation $\varphi_m$.  
The $F$-term contributions are, 
\bea
F_{\Phi_{m}} &\equiv& \frac{\partial W}{\partial \Phi_{m}} \label{ftgen}\,\, . 
\eea
The $\varphi_m$ are (space-time) scalar superpartners     
of the chiral spin-$\half$ fermions $\psi_m$, which together  
form a superfield $\Phi_{m}$.
Since all of the $D$ and $F$ contributions to (\ref{vdef}) 
are positive semidefinite, each must have 
a zero expectation value for supersymmetry to remain unbroken.

For an Abelian gauge group, the $D$-term (\ref{dtgen}) simplifies to
\bea
D^{i}&\equiv& \sum_m  Q^{(i)}_m | \varphi_m |^2 \label{dtab}\,\,  
\eea
where $Q^{(i)}_m$ is the $U(1)_i$ charge of $\varphi_m$.  
When an Abelian symmetry is anomalous, that is,
the trace of its charge 
over the massless fields is non-zero, 
\bea
\Tr Q^{(A)}\ne 0\,\, ,
\label{qtnz}
\eea 
the associated $D$-term acquires a Fayet-Iliopoulos (FI) term,
$\eps\equiv\frac{g^2_s M_P^2}{192\pi^2}\Tr Q^{(A)}$, 
\bea
D^{(A)}&\equiv& \sum_m  Q^{(A)}_m | \varphi_m |^2 
+ \eps \, .
\label{dtaban}  
\eea  
$g_{s}$ is the string coupling and $M_P$ is the reduced Planck mass, 
$M_P\equiv M_{Planck}/\sqrt{8 \pi}\approx 2.4\times 10^{18}$ GeV.
It is always possible to place the total anomaly into a single $U(1)$. 

The FI term breaks supersymmetry near the string scale,
\bea 
V \sim g_{s}^{2} \eps^2\,\, ,\label{veps}
\eea  
unless it can be canceled by a set of scalar VEVs, $\{\vev{\varphi_{m'}}\}$, 
carrying anomalous charges $Q^{(A)}_{m'}$,
\beq
\vev{D^{(A)}}= \sum_{m'} Q^{(A)}_{m'} |\vev{\varphi_{m'}}|^2 
+ \eps  = 0\,\, .
\label{daf}
\eeq
To maintain supersymmetry, a set of anomaly-canceling VEVs must 
simultaneously be $D$-flat 
for all additional Abelian and the non-Abelian gauge groups, 
\beq
\vev{D^{i,\alpha}}= 0\,\, . 
\label{dana}
\eeq
A consistent solution to all (\ref{dana}) constraints 
specifies the overall VEV ``FI-scale'', $\vev{\alpha}$, of
the model.  A typical FNY value is $\vev{\alpha} \approx 7 \times 10^{16}$ GeV.

\section{The group $SU(2)$}
\subsection{The $SU(2)/SO(3)$ Connection\label{suso}}

For the case of $SU(2)$, $T^{SU(2)}_a$ will take on the values of the three Pauli matrices,
\beq
\sigma_x =
\left (                    
\begin{array}{cc}
0 & 1 \\
1 & 0 \\
\end{array} \right ), \,\,
\sigma_y =
\left (
\begin{array}{cc}
0 & -i \\
i &  0 \\
\end{array} \right ), \,\,
\sigma_z =
\left (
\begin{array}{cc}
1 & 0 \\
0 & -1 \\
\end{array} \right ).
\label{pauli}
\eeq
Each component of the vector $\vec{D}$ in this internal space will be the total,
 summed over all fields of the gauge group, ``spin expectation value'' in the given direction.  
Vanishing of the $\vev{\vec{D}\cdot\vec{D}}$ contribution to
$\vev{V}$ demands that $SU(2)$ VEVs be chosen such that the total
$\hat{x}, \hat{y},$ and $\hat{z}$ expectation values are individually zero.
The normalization length, $S^{\dagger}S$, of a ``spinor'' $S$ 
will generally be restricted to integer units by Abelian $D$-flatness constraints from the Cartan
 sub-algebra and any extra $U(1)$ charges carried by the doublet 
(cf.\  Eq.\  \ref{dtab} with $S^{\dagger}S$ 
playing the role of $| \varphi |^2$).
Each spinor then has a length and direction associated with it 
and $D$-flatness requires the sum, placed tip-to-tail, to be zero.
This reflects the generic non-Abelian $D$-flatness 
requirement that the norms of non-Abelian field VEVs are in a one-to-one 
association with a ratio of powers of a corresponding \NA gauge 
invariant \cite{Luty:1995sd}. 

It will be useful to have an explicit (normalized to $1$) representation for 
$S(\theta, \phi)$.  This may be readily obtained by use of the rotation matrix,
\beq
R(\vec{\theta}) \equiv
e^{-i\frac{\vec{\theta}\cdot\vec{\sigma}}{2}} = 
\cos(\frac{\theta}{2}) -i \hat{\theta}\cdot\vec{\sigma}\sin(\frac{\theta}{2})\,\, ,
\label{rotmax}
\eeq
to turn 
$\left (
\begin{array}{c}
1 \\
0 \\
\end{array} \right )$
$\equiv \vert +\hat{z}\rangle$ through an angle $\theta$ about the axis
$\hat{\theta} = - \hat{x} \sin{\phi} + \hat{y} \cos{\phi}$.  The result,
$\left (
\begin{array}{c}
\cos{\frac{\theta}{2}} \\
\sin{\frac{\theta}{2}} \, e^{i \phi} \\
\end{array} \right )$
, is only determined
up to a phase and the choice
\beq
S(\theta, \phi) \equiv
\left ( 
\begin{array}{c}
\cos{\frac{\theta}{2}} \, e^{-i \frac{\phi}{2}} \\
\sin{\frac{\theta}{2}} \, e^{+i \frac{\phi}{2}} \\
\end{array} \right )
\label{spinor}
\eeq
will prove more convenient in what follows.  Within the range of physical 
angles, $\theta = 0 \rightarrow \pi$ and $\phi = 0 \rightarrow 2\pi$, each 
spinor configuration is unique (excepting $\phi$ phase 
freedom for $\theta = 0,\pi$) and
carries a one-to-one geometrical correspondence.  
Up to a complex coefficient,
the most general possible doublet is represented.

A non-trivial superpotential $W$ additionally imposes numerous constraints on allowed
sets of anomaly-canceling VEVs, through the $F$-terms in (\ref{vdef}).
$F$-flatness (and thereby supersymmetry) can be broken through an 
$n^{\rm th}$-order $W$ term containing $\Phi_{m}$ when all of the additional 
fields in the term acquire VEVs,
\bea
\vev{F_{\Phi_m}}&\sim& \vev{{\frac{\partial W}{\partial \Phi_{m}}}} 
      \sim \lambda_n \vev{\varphi}^2 (\frac{\vev{\varphi}}{\MS})^{n-3}\,\, ,
\label{fwnb2}
\eea
where $\varphi$ denotes a generic scalar VEV.
If $\Phi_{m}$ also carries a VEV, then
supersymmetry can be broken simply by $\vev{W} \ne 0$.
For both practical and philosophical reasons, the (Abelian) $D$-condition is usually enforced first.
Unless this constraint holds, supersymmetry will be broken near $M_P$.
On the other hand, the $F$-condition is not all-or-nothing, since the {\it order} of a given dangerous term fixes the scale at which SUSY fails\footnote{The lower the order of an $F$-breaking term, the closer the supersymmetry breaking scale is to the string scale.}.
$F$-flatness must be retained up to an order
in the superpotential that is consistent with observable sector
supersymmetry being maintained down to near the electroweak (EW) scale. 
However, it may in fact be {\it desirable} to allow such a term to escape at some elevated order, since it is known that supersymmetry does {\it not} survive down to `everyday' energies.
Depending on the string coupling strength, $F$-flatness cannot be broken
by terms below eighteenth to twentieth
order\footnote{As coupling strength increases,
so does the required order of flatness.}. 

\subsection{Non-Abelian Flat Directions and Self-Cancellation\label{NASC}}

In \cite{Cleaver:1999cj} we classified MSSM producing singlet field 
flat directions of the FNY model and in \cite{Cleaver:1999mw}
we studied the phenomenological features of these singlet directions.
Our past investigations suggested that for several phenomenological reasons,
including production of viable three generation quark and lepton mass matrices
and Higgs $h$-$\lh$ mixing,
non-Abelian fields must also acquire FI-scale VEVs.

\setcounter{footnote}{0}

In our prior investigations we 
generally demanded ``stringent'' flatness. 
That is, we forced each superpotential term to satisfy $F$-flatness
by assigning no VEV to at least two of the constituent fields.
While the absence of any non-zero terms from within $\vev{F_{\Phi_m}}$ and 
$\vev{W}$ is clearly sufficient to guarantee $F$-flatness along 
a given $D$-flat direction, 
such stringent demands are not necessary.
Total absence of these terms can be relaxed, so long as they appear in
collections which cancel among themselves in 
each $\vev{F_{\Phi_m}}$ and in $\vev{W}$. 
It is desirable to examine the mechanisms of such cancellations
as they can allow additional flexibility
for the tailoring of phenomenologically viable particle properties while
leaving SUSY inviolate.\footnote{Research along this line for the FNY MSSM
is currently underway.}    
It should be noted that success along these lines may be short-lived,
with flatness retained in a given order only to be lost at one slightly higher.

Since Abelian $D$-flatness constraints limit only VEV magnitudes, we are left
with the gauge freedom of each group (phase freedom, in particular, is
ubiquitous) with which to attempt a cancellation between terms (whilst
retaining consistency with non-Abelian $D$-flatness).
However, it can often be the 
case that only a single term from $W$ becomes an offender in a given
$\vev{F_{\Phi_m}}$ (cf.\  Table 1B of \cite{Cleaver:2001ab}).  
If a contraction of \NA fields (bearing
multiple field components) is present it may be possible to effect a
{\it self-cancellation} that is still, in some sense, ``stringently'' flat. 

Near the string scale the complete FNY gauge group is 
\bea
&&[SU(3)_{C}\times SU(2)_{L}\times U(1)_{C}\times U(1)_{L}\times
U(1)_{A}\times \prod_{i=1'}^{5'} U(1)_i\times U(1)_4]_{\rm obs} \times
\nonumber\\
&&[SU(3)_H\times SU(2)_H\times SU(2)_{H'}
\times U(1)_{H}\times U(1)_{7} \times U(1)_{9}]_{\rm hid}\, .
\label{entgg}
\eea
The FNY non-Abelian hidden sector fields are triplets of $SU(3)_H$ or 
doublets of $SU(2)_H$ or $SU(2)_{H'}$. 
Self-cancellation of $F$-terms, that would otherwise break 
observable sector supersymmetry far above the electro-weak scale,
might be possible for flat directions containing such doublet or triplet 
VEVs. 
Since intermediate scale $SU(3)_H$ triplet/anti-triplet condensates 
are more likely to produce viable observable sector electro-weak scale 
supersymmetry breaking than are their $SU(2)_{H^{(')}}$ counterparts,  
we focus herein
on \NA directions containing doublet, but not triplet, FI-scale VEVs.

Whenever ``spinors'' of $SU(2)$ appear in $W$, they are not of the form
$S^{\dagger}S$, but rather are in the antisymmetric contraction
\beq
S_1 \cdot S_2 
\equiv
S^T_1\, 
i \sigma_2\, 
S_2 =
S^T_1
\left (
\begin{array}{cc}
 0 & 1 \\
-1 & 0 \\
\end{array} \right )
S_2 \,\, .
\eeq
This form, which avoids complex conjugation and thus satisfies the requirement
of analyticity, is also rotationally (gauge) invariant as can be verified using
$\{\sigma_i , \sigma_j\} \nolinebreak[4] = \nolinebreak[4] 2 \delta_{ij}$,
$[\sigma_2 , \sigma_2] \nolinebreak[4] = \nolinebreak[4] 0$, 
and Eqs.\ (\ref{pauli}, \ref{rotmax})\, :
\beq
\sigma_2 R(\vec{\theta}) =
R^{\ast}(\vec{\theta}) \sigma_2
\eeq
\beq
S_1' \cdot S_2' =
(RS_1)^T (i \sigma_2) (RS_2) =
S_1^T (i \sigma_2) (R^{\dagger}R) S_2 =
S_1 \cdot S_2\, .
\eeq
From Eq.\ (\ref{spinor}), the general form of such a contraction may
be written explicitly as 
\beq
\label{contract}
S(\theta,\phi) \cdot S(\Theta,\Phi) =
- \sin(\frac{\theta - \Theta}{2})\cos(\frac{\phi - \Phi}{2})
-i \sin(\frac{\theta + \Theta}{2})\sin(\frac{\phi - \Phi}{2})
\,\, .
\eeq
The magnitude of this term must be a purely geometrical quantity and can be
calculated as
\beq
\vert S(\hat{n}) \cdot S(\hat{N}) \vert =
\sqrt{\frac{1 - \hat{n} \cdot \hat{N}}{2}} =
\sin(\frac{\delta}{2}) 
\,\, , 
\eeq
where $\delta (0 \rightarrow \pi)$ is the angle between $\hat{n}$ and $\hat{N}$.
The absence of a similar concise form for the phase is not a failing of
rotational invariance, but merely an artifact of the freedom we had in choosing
(\ref{spinor}).
Self-cancellation of this term is independent of the spinors' lengths and
demands only that their spatial orientations be parallel\footnote{The
contraction of a field with itself vanishes trivially.}. The same conclusion
is reached by noting that anti-symmetrizing the equivalent (or proportional)
spinors yields a null value.
\setcounter{footnote}{0}
VEVs satisfying this condition are clearly not
$D$-consistent unless other \NA VEVed fields also exist such that the
{\it total} ``spin'' vector sum remains zero\footnote{In the notation of
\cite{Cleaver:1999mw}, taking a single sign for each of the $s_{k'}$ is a special case of
\NA self-cancellation, as is $\sum_{k=1}^p n_k s_k = 0$ a special case of the
$D$-constraint.}.  To examine generic cases of cancellation
between multiple terms, the full form of (\ref{contract}) is needed.

As an important special case, consider the example of a superpotential term
$\phi_1 \ldots \phi_n S_1 S_2 S_3 S_4$\footnote{Here and in the following
discussion we consider the doublets of a single symmetry group.} with
$\phi_n$ Abelian.  This is shorthand
for an expansion in the various pairings of non-Abelian fields, 
\beq
W \propto
\phi_1 \ldots \phi_n
\{(S_1 \cdot S_2)(S_3 \cdot S_4) +
(S_2 \cdot S_3)(S_1 \cdot S_4) +
(S_3 \cdot S_1)(S_2 \cdot S_4)\}\, ,
\label{na4}
\eeq
Broadly, we notice that:
\begin{itemize}
\item Whenever each term holds the same field set, the spinors may be treated as
normalized to one, bringing any larger magnitudes out front as overall factors.
Furthermore, since $S^T$ appears but never $S^{\dagger}$, the same can
be done with any phase selections.
\item Since the contractions are antisymmetric, sensible interpretation of terms
with multiple factors demands the specification of an ordering.
\end{itemize}
The appropriate ordering, or equivalently the choice of relative signs, for
(\ref{na4}) is such to ensure {\it total} anti-symmetrization.
When (\ref{na4}) is explicitly evaluated using the previously established
formalism it is seen to vanish identically for {\it all} field values.
The calculation is simplified without loss of generality by taking 
$\theta_1 = \phi_1 = \phi_2 = 0$.  We emphasize the distinction between
this identical exclusion from the superpotential and cancellations which exist
only at the vacuum expectation level.  $W$-terms with $6$ \NA fields are formed
with factors of (\ref{na4}) and also vanish, as do all higher order terms.

Even safe sectors of $W$ (in particular with $\vev{\Phi_{m}} = 0$) may yield
dangerous $\vev{F_{\Phi_{m}}} \nolinebreak[4] \equiv \nolinebreak[4]
\vev{\frac{\partial W}{\partial \Phi_{m}}}$
contributions.  The individual $F$-terms may be separated into two
classes based on whether or not $\Phi_{m}$ is Abelian.  For
the case of $\Phi_{m}$ non-Abelian, $\vev{F_{\Phi_{m}}}$ is itself a doublet.
As a note, terms like $\vev{F_{S_4}} \nolinebreak[4] \equiv \nolinebreak[4]
\vev{\frac{\partial W}{\partial S_4}}$ which {\it would have} arisen out of
(\ref{na4}) are cyclically ordered and also vanish identically.

\section{Minimal Standard Heterotic-String Model Non-Abelian Flat Directions\label{mssmfds}} 

Our initial systematic search for MSSM-producing stringent flat directions 
revealed four singlet directions that were flat to all order, 
one singlet direction flat to twelfth order, and numerous singlet 
directions flat only to seventh order or lower \cite{Cleaver:1999cj}. 
For these directions, renormalizable mass terms appeared for one complete
set of up-, down-, and electron-like fields and their conjugates.
However, the apparent top and bottom quarks did not appear in the 
same $SU(2)_L$ doublet. Effectively, these flat directions gave the
strange quark a heavier mass than the bottom quark. This inverted 
mass effect was a result of the field $\p{12}$ receiving a VEV
in all of the above directions.

We thus performed a search for MSSM-producing singlet flat directions
that did not contain $\vev{\p{12}}$. None were found. 
This, in and of itself, suggests the need for 
non-Abelian VEVs in more phenomenologically
appealing flat directions. 
Too few first and second generation down and electron mass terms
implied similarly. 

Among the FNY MSSM non-Abelian flat directions investigated in 
\cite{Cleaver:2000aa,Cleaver:2001ab}, that denoted FDNA(5+8) generated the best phenomenology.
The search for dangerous terms yielded 131 results, five of them to $\vev{W}$ with
${11}^{th}$ the lowest order and 126 of them to $\vev{F}$, as low as order
four (counting variations of the four fields labeled $(\p4)$ only once).
World sheet selection rules
reduced this number to 32, all of them $F$-terms.  Disallowing more than two
\NA fields (for each $SU(2)$ group)
trimmed the list further to just the eight terms in
Table~\ref{T:danger8}.  If a single incidence of $(\p4)$ is mandated, then
it is so indicated by a lack of parenthesis.

\begin{table}[htb]
\caption{Surviving Candidates for non-Abelian Cancellation}
\begin{center}
\vspace{\baselineskip}
{\begin{tabular}{l|l|l}
\#& {\cal{O}}(W) & $F$-term \\ \hline\hline
1 &  4 & $H^s_{16}\, \vev{\H{26} \cdot \V{37}} \vev{N^c_3}$ \\ \hline
2 &  5 & $V^s_{32}\, \vev{\H{26} \cdot \V{37}} \vev{\p{4} \sH{37}}$
        \\ \hline
3 &  5 & $V_{15}\, \vev{\cdot \V{35}} \vev{\pp{4} \sH{30} \sH{21}}$ \\ \hline
4 &  5 & $V_{17}\, \vev{\cdot \V{5}} \vev {\pp{4} \sH{30} \sH{15}}$ \\ \hline
5 &  8 & $\bp{13}\,  \vev{\H{26} \cdot \V{37}} \vev{(\p{4}) \Hs{31} \sH{30}
        \sH{15} N^c_3}$ \\ \hline
6 &  9 & $\bp{13}\,  \vev{\V{5} \cdot \V{35}} \vev{\p{23} {(\p{4})}^2
        {\sH{30}}^2 \sH{21} \sH{15}}$ \\ \hline
7 &  9 & $\bp{12}\,  \vev{\H{26} \cdot \V{37}} \vev{\p{23} (\p{4}) \Hs{31}
        \sH{30} \sH{15} N^c_3}$ \\ \hline
8 & 10 & $\sH{36}\,  \vev{\H{26} \cdot \V{37}} \vev{\p{23} \p{4} \Hs{31}
        \sH{30} \sH{15} \sH{37} N^c_3}$
\end{tabular}}
\end{center}
\label{T:danger8}
\end{table}

The lowest order potentially dangerous F-term (designated as \#1)
contains a factor of $\vev{\H{26}\cdot \V{37}}$ 
which we would like to cancel, as per the discussion of Section
\ref{NASC}.  This requires the VEV orientations to be chosen parallel
in the three-dimensional ${SU(2)}_H$ adjoint space.
Since FDNA(5+8) contains (two) additional \NA fields with VEVs ($\V{5}$ and
$\V{35}$) which can oppose $\H{26}$ and $\V{37}$ with an equal total magnitude,
this choice is also $D$-consistent.  The same factor appears in
and eliminates additionally dangerous Table~\ref{T:danger8} $F$-terms
\#2,5,7 and 8.  Since the other two \NA VEVs had to be
parallel as well, the contraction $\vev{\V{5} \cdot \V{35}}$ in term \#6 is
also zero.  In the language of \cite{Cleaver:2000aa}, we could have said
$s_{\H{26}} = s_{\V{37}} = 1$ and $s_{\V{5}} = s_{\V{35}} = -1$.
This leaves us with only \#3 and \#4, both of which are fifth order
terms with un-VEVed {\it \NA} fields so that self-cancellation is impossible.
Furthermore, they will appear in different $F$-terms and each allows only
a single $(\p4)$ configuration, 
ruling out a couple of other (less satisfactory)
scenarios.  The choice $\vev{\pp4} = 0$, along with $\vev{\ppb4} = 0$ for 
consistency with Eqs.\ (3.21,3.22) of \cite{Cleaver:2001ab}, would restore
$F$-flatness by simply removing the offending terms from $\vev{F}$.
However,  as was discussed in \cite{Cleaver:2001ab}, 
this seems phenomenologically inviable and so it appears that
we are stuck with a broken FDNA(5+8) at order five.\footnote{As a note, the
cancellations which were successful are insensitive to the factor of 18 between
flat directions FDNA5 and FDNA8. (See Table 1.A of \cite{Cleaver:2000aa}.)} 
Also, while it is common to see the vanishing of terms with excessive \NA doublets, these
mark the {\it only} examples wherein \NA self-cancellation by selected VEVs
was found for the `Table 1A' (of~\cite{Cleaver:2001ab}) flat directions.

\section{The $SO(2n)$ Lie Groups}
\subsection{General Properties}

We will now turn our attention to the general case of $SO(2n)$, the rotation group in an even dimensional space.
Wherever a concrete example is needed, $SO(6)$ will be focused on for the sake of its relevance to 
existing studies \cite{Antoniadis:1987dx,Antoniadis:1987tv,Antoniadis:1988tt,Lopez:1990yk,Lopez:1990wt,Lopez:1991ac,Ellis:1997ni,Ellis:1998nk,Ellis:1999ce,Ellis:1999uq} of a string-derived flipped $SU(5)$ GUT.
It is seen quickly in these cases that certain coincidences and simplicities afforded by the $SU(2)$/$SO(3)$ example are no longer available, and that our techniques have to be adjusted accordingly.
Firstly, the case of rotation in three dimensions is very special.
Any transformation is facilitated by the use of three angles and three generators.
There is a unique axis normal to to any rotation plane and our three rotation generators, which define the adjoint space, can be labeled in one-to-one correspondence with unit vectors from the fundamental space.
It is only in this case that the familiar cross-product can be defined.
Secondly, since all generators fail to commute, we have in $SU(2)$ the simplest case of a rank $1$ group, with only a single diagonalizable matrix.

Since the elements of a transformation between coordinate sets are the projections of unit rows from one basis onto unit columns of another, the transpose of this matrix must interchange these roles and yield the inverse operator.
This property, named orthogonality for the nature of the eigenvectors used to diagonalize symmetric matrices, is the origin of the familiar ``dot product'' rotational invariant.
Orthogonal groups are spanned by antisymmetric generators,
\beq
{(e^M)}^T = e^{M^T} = e^{-M}\, ,
\eeq
and by satisfying the condition that
\beq
{\rm det}(e^M) = e^{\Tr(M)} = e^0 = 1\, ,
\eeq
are also licensed to bear the designation of ``special''.
The $(i,j)^{\rm th}$ element of the rotation generator between Cartesian axes $\hat{a}$ and $\hat{b}$ is determined in the small angle limit:
\beq
{(M^{ab})}_{ij} = {\delta}_i^a{\delta}_j^b - {\delta}_i^b{\delta}_j^a\, .
\label{rotgen}
\eeq
As with all rotation operators, these matrices respect the algebra
\beq
[M^{ab},M^{cd}] = \delta^{ad}M^{bc} + \delta^{bc}M^{ad} - \delta^{ac}M^{bd} - \delta^{bd}M^{ac}\, ,
\label{rotalg}
\eeq
which is specified almost entirely by group closure and symmetries.
We will prefer to multiply each group element by a factor of $i$ and go to a trivially complex  Hermitian representation.

The number of generators required in an $m$ dimensional space is the number of ways to take pairs of axes, ${m\choose 2} \equiv \frac{m!}{2!(m-2)!} = \frac{m(m-1)}{2}$, or equivalently, the number of elements composing a general antisymmetric $m \times m$ matrix.
For the case of $SO(6)$, there are $15$ possible rotation planes, and $15$ corresponding angles needed to specify the general rotation.
Although only five angles are needed to provide the orientation of a $6$-vector, there remains an orthogonal $5$-space, with four angles, which leaves the vector intact.
Perpendicular to this rotation is a $4$-space with three angles, and so emerges another way of counting the $\sum_{\alpha=1}^{m-1}\alpha = \frac{m(m-1)}{2}$ generators needed to enact all possible transformations. 

Certain consequences can be seen immediately when the number of generators exceeds the fundamental group dimension.
First, the conception of a rotation axle must be abandoned in favor of the rotation plane.
Secondly, we see that spanning the adjoint space will require an even {\it larger} basis, and we cannot expect that transformations in the fundamental space will be able to realize generic `expectation value' orientations in the chosen generator set.
Since some regions of the adjoint space may be inaccessible, it will be preferable to reverse the procedure of section (\ref{suso}), and instead carry discussion of the $D$-term into the fundamental space.

The rank of $SO(2n)$, or number of mutually commuting generators, is equal to $n = \frac{m}{2}$, the number of independent rotation planes, or one half the spatial dimension.
Each diagonal matrix has eigenvalues of $(+1, -1)$, with the remaining entries all zero.
The non-diagonal matrices can be combined into raising and lowering operators of unit strength, always acting in two diagonal sectors while leaving the rest unaffected.
Every possible pairing of sectors and choice of raising or lowering in each sector is represented.
Another counting exercise, with the ${n\choose 2}$
ways to pick two diagonal generators, times a factor of $4$ for the choices
$\left (
\begin{footnotesize}
\begin{array}{cccc}
+ & + & - & - \\
+ & - & + & -
\end{array}
\end{footnotesize}
\right )$,
plus $n$ for the diagonal matrices themselves, yields the correct total of $2n(n-1) + n = \frac{m(m-1)}{2}$, and verifies the consistency of this construction.
The commutation relations between the operators obey a sort of charge conservation, with the net raising or lowering weight for each sector preserved across the equality.
For SO(6), these matrices are shown explicitly in both the original and diagonal bases, along with example commutators, in the appendix.

Diagonalization of the secular equation 
${\rm det}(M-\lambda\Io) = 0$ for a matrix $M$, yields a product 
$\prod_{i=1}^m(\lambda_i - \lambda)$, whose roots specify the 
matrix eigenvalues, and whose expansion
\beq
(a_0=1)\lambda^m + a_1\lambda^{m-1} + a_2\lambda^{m-2} + \cdots + a_m = 0
\label{secular}
\eeq
produces $m$ coefficients $a_i$, which must be invariant under group 
transformations\footnote{Any `similarity' transformation which is enacted as $S^{-1}MS$ will preserve these invariants. Orthogonal and unitary mappings are the most notable examples.}, as is the determinant they arise from.
The coefficients are sums of products of eigenvalues,
\eg $a_1 = \sum_{i=1}^m \lambda_i$, $a_2 = \sum_{i>j=1}^m \lambda_i \lambda_j$, $a_3 = \sum_{i>j>k=1}^m \lambda_i \lambda_j \lambda_k$, and $a_m = \prod_{i=1}^m \lambda_i$, which provides concrete verification for the claim of invariance.
These factors, equivalent to a full knowledge of the eigenvalues, fully encode all the rotationally invariant properties of $M$, and fully specify all forms into which $M$ may be rotated.
However, the specific combinations shown prove to have their own desirable properties.
For example $a_1$ and $a_m$ may be recognized as the familiar trace and determinant.
In general, all $m$ coefficients may be constructed as a function only of the matrix $M$, by use of the recursive form
\beq
a_j = \frac{1}{j}\sum_{i=1}^j {(-1)}^{i+1} \Tr (M^i) a_{j-i}\, ,
\label{recursion}
\eeq
referenced to the starting value $a_0 \equiv 1$.
For the generators of $SO(2n)$, the odd-valued coefficients vanish automatically, due to the tracelessness (antisymmetry) of odd powers.
It is important to note that this leaves only $n$ constraints (equal to the group rank) to be actively satisfied in any basis change of the generators.
For $SO(6)$, we have specifically:
\beq
a_2 = \frac{-\Tr M^2}{2}, \,\,
a_4 = \frac{-\Tr M^4 -a_2 \Tr M^2}{4}, \,\,
a_6 = \frac{-\Tr M^6 -a_2 \Tr M^4 -a_4 \Tr M^2}{6}\, .
\eeq

Similar to Eq.\ (\ref{secular}), is a theorem by Euler stating 
that $\prod_{i=1}^m (M - \lambda_i\Io) = 0$, 
as an operation on any of the $m$ $\vert\lambda_i\rangle$ must be null.
This equation may, in principle, be used to reduce by one, with each application, the highest power of a series in the matrix $M$, until the limiting case where that power is $m-1$.
The ability to envision this procedure justifies the statement that the Taylor expansion for a function of $M$ is truncated to order $m-1$.
This knowledge is critical to studying the finite rotation operator
\beq
e^{-i\Theta_{ab}M_{ab}}
\, .
\label{fro}
\eeq

We will focus initially on the case of just a single rotation plane and angle $\theta$ in $6$-space, with the adjoint unit vector $\hat{n}$ providing our combination of generators, $\theta (\vec{M}\cdot\hat{n})$.
An explicit representation, consistent with Eq.\ 
(\ref{rotgen}), of the generic planar generator can be formed readily, where $\hat{a}$ and $\hat{b}$ are orthogonal unit $6$-vectors and the notation $\otimes$ is defined by the construction of a tensor, ``$\vert\;\rangle\langle\;\vert$'', out of the column to its left and the row to its right.
\beq
\vec{M}\cdot\hat{n}
\Leftrightarrow
M_{\hat{a}\hat{b}}
\equiv
i(\hat{a}\otimes\hat{b} - \hat{b}\otimes\hat{a})
\label{vectgen}
\eeq
It is clear from this construction that all single plane generators obey the rule
\beq
{(\vec{M}\cdot\hat{n})}^2
\Leftrightarrow M_{\hat{a}\hat{b}}
\cdot
M_{\hat{a}\hat{b}}
=
(\hat{a}\otimes\hat{a} + \hat{b}\otimes\hat{b})\, ,
\label{spg}
\eeq 
meaning that ${(\vec{M}\cdot\hat{n})}^3 = \vec{M}\cdot\hat{n}$, and ${(\vec{M}\cdot\hat{n})}^4 = {(\vec{M}\cdot\hat{n})}^2$, etc. and thus that the rotation operator reduces further in this case, to only 3 terms with 3 undetermined coefficients $(\alpha,\beta,\gamma)$.
\beq
e^{-i\theta (\vec{M}\cdot\hat{n})} =
\alpha + \beta (\vec{M}\cdot\hat{n}) + \gamma {(\vec{M}\cdot\hat{n})}^2
\label{rotopunfixed}
\eeq
Not coincidentally, this is also the number of available discrete eigenvalues, and forcing consistency in Eq.\ (\ref{rotopunfixed}) when $\vec{M}\cdot\hat{n}$ is replaced by $(+1, 0, -1)$, allows us to fix the parameters shown:
\beq
e^{-i\theta (\vec{M}\cdot\hat{n})} =
\Io - i \sin\theta (\vec{M}\cdot\hat{n}) + (\cos\theta - 1) {(\vec{M}\cdot\hat{n})}^2\, .
\label{rotop}
\eeq

With this reduced case in hand, we can now return attention to the general finite $SO(2n)$ transformation specified by the linear combination of rotation generators $\Theta_{ab}M_{ab}$.
To start, we will write the prototype block diagonalized form of a member of this class with $n$ free angles $\theta_i$. 
\beq
i \times
\left (
\begin{footnotesize}
\begin{array}{ccc}
\left(\begin{array}{cc}0&\theta_1\\-\theta_1&0\end{array}\right) &   & \\
  & \left(\begin{array}{cc}0&\theta_2\\-\theta_2&0\end{array}\right) & \\
  &   & \ddots
\end{array}
\end{footnotesize}
\right )
\label{rotangles}
\eeq
Just as the operator of Eq. (\ref{fro}) enacts changes of basis on real $2n$-vector states, {\it orthogonal pairs} of such operators invoke the corresponding similarity (rank-$2$ tensor) transformation on {\it elements of} $SO(2n)$ {\it itself}\footnote{This ensures then that the same result is achieved whether the change of basis occurs before or after operation of the $SO(2n)$ element on a given vector.}, \ie specific {\it instances of} Eq. (\ref{fro}).
In the Taylor sense, the same transformation is then applied to the selected generators, transforming them into the alternate basis\footnote{Or alternatively, performing an opposing rotation while the basis stays fixed, depending on taste.}.
\beq
{(\Theta_{ab}M_{ab})}^{\prime} \equiv
e^{-i\Phi_{cd}M_{cd}}
(\Theta_{ab}M_{ab})
e^{+i\Phi_{cd}M_{cd}}
\label{genprime}
\eeq

Rotation can only map a linear sum of $SO(2n)$ generators into {\it another} such linear sum.
Preservation of the $m$ rotational invariants is the only restriction on what members of this class  may be interrelated by operation of $SO(2n)$.
As is necessarily expected, orthogonal transformations explicitly protect the property of (anti)symmetry, ensuring that the odd invariants remain identically zero.
\beq
{(M^{\prime})}^T \equiv {(OMO^{-1})}^T = OM^TO^T = (-)M^{\prime}
\label{syminv}
\eeq
This leaves then only $n$ non-trivial constraints, which should always be absorbable via some action of the group {\it itself}, into the $n$ angles of Eq. (\ref{rotangles}).
Specifically, for the case of $m=6$, we have:
\bea
a_2 &=& -(\theta_1^2 + \theta_2^2 + \theta_3^2) \nonumber \\
a_4 &=& ((\theta_1 \theta_2)^2 + (\theta_2 \theta_3)^2 + (\theta_3 \theta_1)^2) \label{asix} \\
a_6 &=& -(\theta_1 \theta_2 \theta_3)^2\nonumber 
\eea
The solubility\footnote{Mathematica can readily invert the example system. However, the general solutions are quite clumsy in appearance.} of equations like the above is identical to the statement that any matrix $\Theta_{ab}M_{ab}$ may in principle be converted to the desired block-diagonal form under operation of $SO(2n)$.
Note that in each term angles ever appear individually only as squares.

Since the commutativity of these $n$ sectors from (\ref{rotangles}) will be maintained under any group action, it may be inferred that all combinations of generators are a sum over $n$ orthogonal rotation planes.
Thus, the finite rotation operator in (\ref{fro}) can be factored into separate exponential terms containing each distinct plane, without any complications of the Baker--Hausdorff variety.
It is then seen that the general transformation is enacted by a {\it product} of $n$ (\ref{rotop}) copies.

Note that this {\it does not} imply that $SO(2n)$ is equivalent to the factored product $O(2)^n$.
Rather, it simply says that for {\it each} rotation in $SO(2n)$ which you would like to perform there is in principle an identical representation for that {\it particular} operation in terms of a product of planar rotations.

\subsection{Viewing $D$-Terms From the Fundamental Space\label{dview}}

Having understood the $SO(2n)$ group structure to some degree, the next matter which we may wish to consider is the manner in which field VEVs of the fundamental space are transformed into $D$-terms of the adjoint space.
Ultimately we will also be interested in how the concisely realized adjoint $D$-constraint, namely that the sum of all vector contributions be null, is reflected back onto fundamental states. 
It will be useful first to determine the eigenvectors corresponding to generators of rotation in a given plane, such as appear in (\ref{vectgen}).
Within an arbitrary phase, the normalized solutions for the $(+1, -1)$ eigenvalues of $M_{\hat{a}\hat{b}}$ may be written respectively as:
\beq
\ket{+}_{\hat{a}\hat{b}} = (\hat{a} - i\hat{b})/\sqrt{2}\, , \,\,\,
\ket{-}_{\hat{a}\hat{b}} = (\hat{a} + i\hat{b})/\sqrt{2}\, .
\label{pmeigenvects}
\eeq

As expected for a hermitian matrix, these states are mutually orthogonal (${\langle+|-\rangle}_{\hat{a}\hat{b}} = 0$), and may also be chosen with null projections onto the $2n-2$ additional states required to complete the basis.
They are not generally orthogonal to the eigenvectors of overlapping generators, such as ``$M_{\at{a}\hat{c}}$'', which induces rotation in a plane passing through $\hat{a}$ and some vector $\hat{c}$, where $\hat{b}\cdot\hat{c} = 0$. 
It is interesting to note that the real generator basis as established in (\ref{rotgen}) produces intrinsically complex eigenvectors, while the diagonal matrix set described in the Appendix is instead itself complex with the possibility of real eigenvectors.
These two formulations are bridged by a unitary transformation rather than any $SO(2n)$ element.

The `expectation value' of an eigenstate contracted on its corresponding matrix generator is given by:
\beq
\langle+|M_{\hat{a}\hat{b}}|+\rangle \equiv
\frac{i}{2}(\hat{a} + i\hat{b}) \cdot
(\hat{a}\otimes\hat{b} - \hat{b}\otimes\hat{a}) \cdot
(\hat{a} - i\hat{b}) = 1 \, .
\eeq

The same calculation performed with the $\ket{-}_{\hat{a}\hat{b}}$ state will yield $-1$.
We can argue that all other contractions employing the $\ket{\pm}_{\hat{a}\hat{b}}$ are vanishing since the other $n-1$ `diagonal' generators work in orthogonal spatial sections, and the $2n(n-1)$ `raising and lowering' operators cannot bridge an eigenstate to itself.
Thus, the complete $2n(2n-1)$ element $D$-term which emerges out of this state can be determined:
\beq
\ket{\frac{\hat{a} - i\hat{b}}{\sqrt{2}}}
\, \Longrightarrow \,
(\hat{a}\otimes\hat{b} - \hat{b}\otimes\hat{a})_{ij} \, .
\label{dtermplus}
\eeq

Notice that while previously in (\eg~Eq.~\ref{vectgen}) a similar notation was used to express the matrix elements of a {\it single} generator $M_{\hat{a}\hat{b}}$, the same form now provides the `expectation value' result for {\it every} generator, interpreted as either an adjoint vector ($i>j$), or as a fundamental antisymmetric matrix.
For example, if we take $\hat{a} = (1,0,0,\cdots)$ and $\hat{b} = (0,1,0,\cdots)$, then (\ref{dtermplus}) says that the related complex $\mathbf 2n$ dimensional vector will produce a $D$-term contribution of $\vev{M_{12}} = 1 (= - \vev{M_{21}})$, with all other elements zero. 
Covariance suggests that the same expression must also hold for vectors ($\hat{a}, \hat{b}$) which do not lie along a single direction of the selected basis.

We may next wish to inquire what form the $D$-term matrix would take for a fully arbitrary $\mathbf 2n$-plet of VEVs, {\it i.e.} with the constraints $\hat{a}\cdot\hat{a} = \hat{b}\cdot\hat{b} = 1$ and $\hat{a}\cdot\hat{b} = 0$ relaxed.
To avoid confusion with the previous results, we will now refer to the unit vectors $\hat{c}$ and $\hat{d}$, used to compose our general state as:
\beq
\ket{v} \equiv R\hat{c} + i I\hat{d} \, .
\label{ketv}
\eeq

We can still insist without any loss of generality that the state be normalized to unity overall, and in fact this condition facilitates a simple rescaling to integral multiples of the squared FI scale $\mvev{\alpha}$ as is typically required.
With this condition in place, the real coefficients $R$ and $I$ are restricted to $R^2 + I^2 = 1$.

In correspondence to the earlier (\ref{pmeigenvects}), we can also define the states
\beq
\ket{\pm}_{\hat{c}\hat{d}} \equiv (\hat{c} \mp i\hat{d})/\sqrt{2}
\, ,
\label{pmcd}
\eeq

although a word of caution is in order to the effect that {\it no} eigenvalue relation to some matrix $M_{\hat{c}\hat{d}}$ is being posited.
In fact, it is no longer even necessarily true that ${\langle+|-\rangle}_{\hat{c}\hat{d}} = 0$
However, in terms of these states there is an alternate formulation of the general complex state:
\bea
\ket{v} = A\ket{+}_{\hat{c}\hat{d}} + B\ket{-}_{\hat{c}\hat{d}}\, ,
\label{ABvec} \\
R = \frac{A+B}{\sqrt{2}}\,\,,\,\,
I = -\frac{A-B}{\sqrt{2}}\, .
\label{IofA}
\eea 

The normalization condition is here realized as $A^2 + B^2 = 1$.
With these conventions in place the expectation value on any generator $M$ can now be decomposed.
\beq
\langle v|M|v \rangle =
R^2 \langle c|M|c \rangle +
I^2 \langle d|M|d \rangle +
iRI(\langle c|M|d \rangle
 -  \langle d|M|c \rangle)
\label{decomp}
\eeq

We will now focus on the term $\langle c|M|c \rangle$.
Since this is a scalar contraction, the transpose operator will be an identity.
Furthermore, since the vector $\hat{c}$ is real by definition, the state $\langle c|$ is merely ${\ket{c}}^T$, so that $\langle c|M|c \rangle = \langle c|M^T|c \rangle$.
However, the antisymmetry of $M$ says that this expression is equal to its own negative, and must therefore vanish. 
The same result holds for $\langle d|M|d \rangle$, and in fact for {\it any} $D$-term constructed from a purely real vector\footnote{This conclusion is also true for `trivially complex' vectors, as the inclusion of an {\it overall} complex phase can be absorbed in the contraction without altering the remaining discussion.}.
In a sign of consistency, the remaining two terms are proportional to ``$RI$'', such that all $D$-term contributions vanish unless $\ket{v}$ contains {\it both} real and imaginary contributions.
An argument similar to that just given on the transpose suggests that these terms are opposites.
Also, since they are clearly related by complex conjugation, this means that each term is purely imaginary, preserving the necessary reality condition on the expectation value. 
Combining the information above with Eqs.\ (\ref{pmcd},\,\ref{IofA}) allows us to state that:
\beq
\langle v|M|v \rangle =
2iRI \langle c|M|d \rangle =
(A^2 - B^2) \langle +_{\hat{c}\hat{d}}|M|+_{\hat{c}\hat{d}} \rangle
\, .
\label{irasquared}
\eeq 

The normalization imposed on the $A$ and $B$ coefficients restricts the pre-factor $A^2 - B^2$ to range between (-1, 1).
The lesson to be taken from these exercises is that an unbalanced mixing between the `$I$' and `$R$' coefficients of a given fundamental state will not effect the {\it orientation} of the adjoint space $D$-term which it yields, but it can reduce the {\it scale} of that contribution. 

However, to complete analysis of the general $\mathbf 2n$ vector, there remains the matter of $\hat{c}\cdot\hat{d} \neq 0$ to deal with.
In order to proceed, let us introduce an alternate unit vector $\hat{d'}$ which is perpendicular to $\hat{c}$ and lies in the plane defined by $\hat{c}$ and the original $\hat{d}$.
Breaking $\hat{d}$ down into its new basis components, we have:
\beq
\hat{d} = \hat{d'}\sin{\theta} + \hat{c}\cos{\theta} \, ,
\label{dorig}
\eeq
where $\theta$ is the angle separating $\hat{c}$ and $\hat{d}$.

Under this transformation, the state $\ket{v}$ from (\ref{ketv}) becomes
\bea
\ket{v} &=& R\hat{c} + i I(\hat{d'}\sin{\theta} + \hat{c}\cos{\theta})
\label{ketprimeri} \\
&=& \frac{R + iIe^{i\theta}}{\sqrt{2}} {\ket{+}}_{\hat{c}\hat{d'}} \, + \,
\frac{R + iIe^{-i\theta}}{\sqrt{2}} {\ket{-}}_{\hat{c}\hat{d'}} \, .
\label{ketprime}
\eea

\setcounter{footnote}{0}
The main points to notice here are that we can decompose $\ket{v}$ into the orthogonal eigenstates of the pure generator $M_{\hat{c}\hat{d'}}$, and that no single raising or lowering operator can hope to join these states\footnote{The ${\ket{\pm}}_{\hat{c}\hat{d'}}$ are separated by a third `rung' corresponding to some state with a null response to the ($\hat{c}$,$\hat{d'}$) plane rotation operator.}.
Because of this, the resultant $D$-term will again be fully along the $M_{\hat{c}\hat{d'}}$ orientation, and we have only to find its magnitude.
Since the ${\ket{\pm}}_{\hat{c}\hat{d'}}$ coefficients in (\ref{ketprime}) are complex, the expression does not form a realization of (\ref{ABvec}), nor should the results following that equation be expected to apply\footnote{Although (\ref{ABvec}) did in fact represent the {\it general} $\mathbf 2n$-plet, this was for $\hat{c}$ and $\hat{d}$ unconstrained.  Since a definite condition between $\hat{c}$ and $\hat{d'}$ has now been imposed, complex coefficients are required for the general $\theta$.}.

Rather than proceeding directly with (\ref{ketprime}), let us instead evaluate $\langle c|M_{\hat{c}\hat{d'}}|d \rangle = i \langle +_{\hat{c}\hat{d}}|M_{\hat{c}\hat{d'}}|+_{\hat{c}\hat{d}} \rangle$, as appears in (\ref{irasquared}). 
Rewriting (\ref{dorig}) as 
\beq
\hat{d'} = \frac{\hat{d} - \hat{c}\cos{\theta}}{\sin{\theta}}\, ,
\label{dprime}
\eeq
it is readily confirmed that:
\beq
M_{\hat{c}\hat{d'}} \equiv
i(\hat{c}\otimes\hat{d'} - \hat{d'}\otimes\hat{c}) =
i\frac{(\hat{c}\otimes\hat{d} - \hat{d}\otimes\hat{c})}{\sin{\theta}}\, .
\label{mprime}
\eeq

The desired contraction then becomes
\beq
\langle c|M_{\hat{c}\hat{d'}}|d \rangle =
i\frac{\hat{c}\cdot(\hat{c}\otimes\hat{d} - \hat{d}\otimes\hat{c})\cdot\hat{d}}{\sin{\theta}} =
i\frac{1-{\cos}^2(\theta)}{\sin{\theta}} =
i \sin{\theta}\, .
\label{contractcd}
\eeq

Combining this result with (\ref{irasquared}) allows us to summarize the magnitude of the general $D$-term as
\beq
\ket{v} \equiv R\hat{c} + i I\hat{d}
\, \Longrightarrow \,
|D_v| = -2RI\sin{\theta} = (A^2 - B^2)\sin{\theta}\, .
\label{Dmag}
\eeq
where $\theta$ is the angle between $\hat{c}$ and $\hat{d}$, and the relation appearing in (\ref{ABvec}) is also in use.
Extending the intuition that only non-trivially complex states correspond to $D$-terms, the $\sin(\theta)$ term will kill $|D_v|$ if $\hat{c}$ and $\hat{d}$ are proportional, such that $\ket{v}$ is real times an overall phase.
Also, we can note that this scale is exactly sufficient when integrated with the known adjoint direction of (\ref{mprime}) to exactly mimic the simple result of (\ref{dtermplus})~\footnote{Recall, as per the discussion following this equation, that the {\it orientation} corresponding to a state which strikes only a single generator may be written in matrix form as the generator {\it itself}, dropping the factor of $i$.} for the total $D$-term.
Specifically, under the redefinitions $\vec{e} \equiv \sqrt{2}R\hat{c}$ and $\vec{f} \equiv -\sqrt{2}R\hat{d}$, the general result becomes
\beq
\ket{v} \equiv
\ket{\frac{\vec{e} - i\vec{f}}{\sqrt{2}}}
\, \Longrightarrow \, 
(\vec{e}\otimes\vec{f} - \vec{f}\otimes\vec{e})_{ij} \, .
\label{dtermgen}
\eeq

Basic manipulations are enough to reveal an additional equivalent construct:
\beq
(D_v)_{ij} \equiv \langle v|M_{ij}|v \rangle = 
(\frac{|v\rangle\langle v| - |v^* \rangle\langle v^*|}{i})_{ij} \, .
\label{minuscc}
\eeq

Finally, we come to the interesting realization that, having exhausted all available generality in our state $\ket{v}$, and receiving still for this effort only $D$-terms which correspond to single-plane rotation generators as in (\ref{dtermgen}), it is {\it not possible} to represent an arbitrary adjoint space direction by the VEVs of a solitary $\mathbf 2n$-plet.
This conclusion is born out by analogy to the discussion around (\ref{rotangles}), where it was shown that the general $SO(2n)$ transformation matrix, {\it i.e.} adjoint space vector,  required contributions from $n$ orthogonal planar type generators times scaling angles. 

\subsection{An Example of $D$-Flatness in $SO(6)$}

To be more concrete, let us specifically consider $D$-flat directions formed from 
the VEVs of fundamental vector $\mathbf 6$ representations for the gauge group $SO(6)$, 
choosing to express the generators in their `original' basis as shown in the Appendix.
We will denote the (generally complex) VEVs of the $k^{\rm th}$ 6-plet according to,
\bea
< {\mathbf 6}_k> = \left(
\begin{array}{c}
	 \alpha_{k,1} +  i\,  \beta_{k,1} \\
	 \alpha_{k,2} +  i\,  \beta_{k,2} \\
	 \alpha_{k,3} +  i\,  \beta_{k,3} \\
	 \alpha_{k,4} +  i\,  \beta_{k,4} \\
	 \alpha_{k,5} +  i\,  \beta_{k,5} \\
	 \alpha_{k,6} +  i\,  \beta_{k,6} 
\end{array}
\right),
\eea 
where $\alpha_{k,j}$ \&  $\beta_{k,j}$ are real. 
$D$-flatness constraints when applied to a single 6-plet (with field subscript $k=1$)
demand the vanishing of:
\beq
D_{ij} \equiv
(\alpha_{1,a} +  i\,  \beta_{1,a}) \cdot
\{ {\delta}_i^a{\delta}_j^b - {\delta}_i^b{\delta}_j^a \}
\cdot (\alpha_{1,b} +  i\,  \beta_{1,b})
\, ,
\label{directD}
\eeq
as can be enforced by direct computation from (\ref{rotgen}).
Contracting indices for a more concisely worded expression,
\bea
\{ \alpha_{1,i} \beta_{1,j} - \alpha_{1,j} \beta_{1,i} = 0 \}
\, ,
\label{ndall}
\eea
for $i,j=1$ to 6 and $i<j$.
This is of course equivalent to a vanishing of the construct from (\ref{dtermgen}).

The only solutions to (\ref{ndall}) are fundamentally trivial.
Since real and imaginary VEV component coefficients, $\alpha$ and $\beta$, always appear together as products, a solution exists for any pure real or pure imaginary vector.
However, even for those solutions which employ cancellation of $D$-term contributions between 
$\alpha_{1,i}\, \beta_{1,j}$ products, 
$D$-flatness is maintained if and only
if the ratio $\alpha_{1,j}/\beta_{1,j}$ is the same for all non-vanishing members $j$. 
Furthermore,   
neither $\alpha_{1,j}\ne 0$ \& $\beta_{1,j}=0$ nor $\alpha_{1,j}\ne 0$ \& $\beta_{1,j}=0$ is allowed for any $j$. 
This statement of constant phase is in agreement with the statements on `reality conditions' from the previous section, specifically as appears following~(\ref{decomp},~\ref{Dmag})\,. 

It is somewhat more interesting to generalize the above solutions for a single $<\mathbf 6>$ to
a set of $n$ distinct $\mathbf 6$'s, each having respective real and imaginary 
VEV components $\alpha_{k,j} + i\,  \beta_{k,j}$ for $k= 1$ to $n$. 
The constraints become a sum over expressions of the sort in (\ref{ndall}), 
\bea
\{\sum_{k=1}^n \alpha_{k,i} \beta_{k,j} - \alpha_{k,j} \beta_{k,i} = 0 \}
\, ,
\label{neall}
\eea
for $i,j=1$ to 6 and $i<j$.
Rather than forcing reality on just one state, we instead now have a dispersement of the burden of $D$-flatness among every constituent field, none of which need individually follow the former condition.
Each new 6-plet VEV generates further non-trivial solution classes, allowing new possibilities for cancellations between different fields, and additional freedoms for each individual state.
The simplest non-trivial example involves two fields.
If each field is further restricted to non-zero VEVs for only its top two complex components, that is, for $i,j=1$, $2$, then the $D-$flat constraints (\ref{neall}) reduce to simply      
\bea
\alpha_{1,1} \beta_{1,2} - \alpha_{1,2} \beta_{1,1} + \alpha_{2,1} \beta_{2,2} - \alpha_{2,2} \beta_{2,1} = 0\, . \label{ne1}
\eea
Hence we are free to choose seven of the eight VEV components. Some choices
correspond to products of single VEV solutions (\ref{ndall}), while others do not.
The new, non-product solutions are those for which $\alpha_{1,1}/ \beta_{1,1} \ne  \alpha_{1,2}/ \beta_{1,2}$.
One such example is
\bea
\alpha_{1,1} &=& 1 ;\,\, \beta_{1,1}=  2;\,\,
\alpha_{1,2} = 3;\,\, \beta_{1,2} = 5\, ; \nonumber\\
\alpha_{2,1} &=& 7 ;\,\, \beta_{2,1}= 11;\,\,
\alpha_{2,2} = 13;\,\, \beta_{2,2} = 20\, \frac{4}{7}\, . \label{nf1}
\eea

\subsection{Reduction and Interpretation of the Multiple-VEV Constraint}

The condition written in (\ref{neall}) is clearly a full specification of the desired $D$-term result, but it cannot fulfill the {\it spirit} of our search, in that the expression is neither geometrical nor intuitively comprehensible. 
In fact, the number of conditions enforced grows like the group adjoint dimension, and quickly becomes so large that even inelegant and forceful approaches may find the task of its solution insuperable. 
We would much prefer a statement of principle from which one could deduce at a glance the {\it existence} of solutions, followed in short order by {\it specific} field values obedient to the criterion.
Furthermore, the prospect of a condition whose complexity grows {\it linearly} with the group rank $n$, rather than {\it quadratically} is enticingly motivated by the earlier observation that any adjoint vector can be decomposed into $n$ commuting antisymmetric fundamental matrices of the special rotation generator form, (\ref{vectgen}). 
This section will then pursue results useful for understanding $D$-flatness in the presence of multiple $SO(2n)$ VEVs.


We can proceed by denoting each contributing state as $\ket{\gamma}$, so that the net $D$-term will be:
\beq
D_{ij} \equiv
\sum_{\gamma} \langle \gamma | M_{ij} | \gamma \rangle\, .
\label{netD}
\eeq
As discussed around (\ref{rotangles}), this matrix can be viewed as the sum of $n$ orthogonal single-plane rotation matrices times scale factors.
We know from (\ref{dtermplus}) that any such matrix can be composed by the contraction of its `positive' eigenvector around $M_{ij}$.
Furthermore, the independence of each of the $n$ matrices tells us that their eigenvectors will also be eigenvectors of the total $D$-term matrix in (\ref{netD}), which will be written as $\ket{\lambda}$. 
Therefore, we can state the following relationships between the $D$-matrix and its eigenvectors:
\beq
\sum_{\gamma} \langle \gamma | M_{ij} | \gamma \rangle\, =
\sum_{k=1}^n \lambda_k^+ \langle \lambda_k^+ | M_{ij} | \lambda_k^+ \rangle\, =
\sum_{k=1}^{2n} \lambda_k (| \lambda_k \rangle \langle \lambda_k |)_{ij}\, .
\label{netDeq}
\eeq
The front factor of $\lambda_k^+$ is needed in the central expression to represent the scale factor, which shifts the eigenvalue away from unity, of each `planar' member in the composite $D$-term\footnote{In other words, the contraction of $\ket{\lambda}$ around $M_{ij}$ reconstructs the given constituent of $D_{ij}$ with an overall unit scale, but since $\ket{\lambda}$ {\it is} the eigenvector of the matrix to be rebuilt, its eigenvalue $\lambda$ and the desired scale are in fact one and the same.}.
Since $M_{ij}$ is Hermitian and complex, we can see that its eigenvalues always come in $\pm \lambda$ pairs, with eigenvectors related by complex conjugation.
\beq
M|\lambda \rangle = \lambda|\lambda \rangle\,
\Longrightarrow M|\lambda^* \rangle = (M^*|\lambda \rangle)^* =
-(M|\lambda \rangle)^* = -\lambda|\lambda^* \rangle\, .
\label{pmpairs}
\eeq
Also noting that
\beq
\langle \gamma | M_{ij} | \gamma \rangle\, =
- \langle \gamma^* | M_{ij} | \gamma^* \rangle\, .
\label{ccgamma}
\eeq
for any $|\gamma \rangle$, it is clear that the plus sign on $\lambda_k^+$ should really just be taken to indicate that half of the eigenvalues (one from each pair) are in play here, with a symmetry protecting the choice of either the `positive' or `negative' member.
This is in accord with the properties previously observed directly in section (\ref{dview}).
The third expression in (\ref{netDeq}) can be readily justified from the second, via comparison to Eqs.\ (\ref{dtermgen}, \ref{minuscc})\,, when written as:
\beq
\sum_{k=1}^n \lambda_k^+ (| \lambda_k^+ \rangle \langle \lambda_k^+ |
	- | \lambda_k^* \rangle \langle \lambda_k^* |)_{ij}\, =
\sum_{k=1}^n \lambda_k^+ (\hat{a_k}\otimes\hat{b_k} - \hat{b_k}\otimes\hat{a_k})_{ij}\, ,
\label{termsofab}
\eeq
where $| \lambda_k^+ \rangle \equiv (\hat{a_k} -i \hat{b_k})/\sqrt{2}$.
This alternate method of writing the same sum of matrices should be recognized as simply a diagonalization by similarity transformation\footnote{By starting with the conception of a diagonal form, one could in fact read this argument in reverse as an alternate proof of the decomposition into $n$ planes.}.

We can now read directly from (\ref{netDeq}) that the $n$ vectors represented as $\sqrt{\lambda}|\lambda \rangle$ function as a completely equivalent set of input to the original fully general $\ket{\gamma}$'s.
This reinforces the notion that it may be possible to impose some reduced number of conditions equivalent to the rank which will serve to eliminate all $n(2n-1)$ elements of the $D$-matrix.
The only flaw in this approach is our complete inability to know $|\lambda \rangle$ {\it before} we have {\it already} specified the overall VEV set!

From another perspective, we might imagine that a concrete expression for the eigenvalues $\ket{\lambda}$ of the matrix from (\ref{netD}) would enable a clear view of what conditions ensure that these numbers will vanish.
By inserting dual complete sets, each written as $\Io \equiv \sum |\lambda \rangle \langle \lambda|$, into the first expression of (\ref{netDeq}), we can say:
\beq
\sum_{k,l= 1}^{2n}
\langle \lambda_k | M_{ij} | {\lambda_l}' \rangle
\left( \sum_{\gamma}
\langle \gamma | \lambda_k \rangle \langle {\lambda_l}' | \gamma \rangle
\right) =
\sum_{k=1}^n \lambda_k^+ \langle \lambda_k^+ | M_{ij} | \lambda_k^+ \rangle
\, .
\label{complete}
\eeq 

The strength of these equations in the $(i,j)$ is enough to let us equate coefficients of each contraction across the two sides.
This is seen most cleanly in a basis where $M_{ij}$ is diagonal, such that `cross-terms' on the left-hand side of 
(\ref{complete}) vanish\footnote{We are referring here to the $n$ diagonal matrices; the remaining raising and lowering 
operators can join unmatched kets, but the sum of all such terms must vanish since these elements make no contribution on the left.}, and we are left with
\beq
\sum_{\gamma} \left(
{\vert \langle \gamma | \lambda^+ \rangle \vert}^2 - 
{\vert \langle \gamma | \lambda^* \rangle \vert}^2
\right) = \lambda^+
\, ,
\label{lambdaeq}
\eeq 
also using (\ref{ccgamma}).
Depending on your preference, this expression may be regrouped into various forms with distinct interpretations. 
\beq
\lambda =
\sum_{\gamma} \langle \gamma \vert \left(
| \lambda \rangle \langle \lambda | -
| \lambda^* \rangle \langle \lambda^* |
\right) \vert \gamma \rangle =
\langle \lambda \vert
\sum_{\gamma} \left(
| \gamma \rangle \langle \gamma | -
| \gamma^* \rangle \langle \gamma^* |
\right) \vert \lambda \rangle
\, ,
\label{twoforms}
\eeq
The first of these constructions appears as an element of (\ref{netD}), and again validates the notion that the entire matrix could be set to zero via consideration of only a reduced subset of $n$ `diagonal' generators {\it if only} one could know ahead of time which diagonal set to choose!
We can also see that each eigenvalue of the overall $D$-matrix vanishes in turn when its corresponding eigenvector is `real', up to an overall phase.
This is natural in light of the discussion above where it was noted that the $\sqrt{\lambda}|\lambda \rangle$ serve as equivalent input to the VEVed $| \gamma \rangle$'s\footnote{As pointed out around (\ref{Dmag}), real SO(2n) states create no $D$-term contributions.}, and that their eigenvalues flip sign under complex conjugation.
Another interpretation of this expression holds that $\lambda$ is the imbalance between projections onto the `positive' and `negative' eigenvectors. 
The second formulation of (\ref{twoforms}) is arrived at by applying the (free) operation of complex conjugation to the (real) second element of (\ref{lambdaeq}) before separating out the terms.
From this, we can read another condition, applied now to the more tangible input states, which will also kill the $D$-term:
\beq
\sum_{\gamma} | \gamma \rangle \langle \gamma |
\, \Longrightarrow \, 
\textrm{\small{REAL}}
\, .
\label{gammareal}
\eeq
By way of first analysis, we can note that this expression is immune to overall phase factors associated with $\ket{\gamma}$, and that it properly reduces to the `generalized reality' condition imposed on just a single state.
In fact though, the statement of (\ref{gammareal}) contains only the same information, and the same shortcomings of the previous attempts.
Using (\ref{minuscc}), the entire $D$-term can be written
\beq
D_{ij} =
\sum{\gamma} (|\gamma\rangle\langle\gamma| - |\gamma^*\rangle\langle\gamma^*|)_{ij}
\, ,
\label{wholed}
\eeq 
and (\ref{gammareal}) is simply the statement that the matrix should vanish\footnote{The embedding of (\ref{wholed}) within the second expression of $\lambda$ in (\ref{twoforms}) is a consistency check, since the contraction of $\ket{\lambda}$ about the matrix which they represent must identically yield the eigenvalue.}.
\setcounter{footnote}{0}
Alternatively, using the language of (\ref{dtermgen}), we can say that
\beq
\sum_{\gamma} (\vec{e}_{\gamma}\otimes\vec{f}_{\gamma}
- \vec{f}_{\gamma}\otimes\vec{e}_{\gamma})_{ij} = 0\, ,
\label{proj2}
\eeq
where $| \gamma \rangle \equiv (\vec{e}_{\gamma} -i \vec{f}_{\gamma})/\sqrt{2}$.
However, this only serves to mimic (\ref{neall}) and its associated large number of conditions for projections in each of 
the generator planes.

A fully geometric interpretation of non-Abelian $SO(2n)$ $D$-flat directions should offer proper, concise 
criterion with physically intuitive interpretation. 
Along this line, we have made several arguments for the feasibility of solutions 
that grow only like the group rank for increasing number of fundamental {\bf 2n} field VEVs. 
Realization of these arguments, nevertheless, remains an open issue, for after a detailed study, a more 
functional re-expression of (\ref{neall}) has not been found.  



\subsection{Ensuring Simultaneous $SO(2n)$ $F$-Flatness}

For the previous study of $SU(2)$, computations were performed in the adjoint space such that the $D$-condition was 
readily realized, while effort was required to transfer $F$-terms into the language of orientations in this space 
(cf.~Eq.~\ref{contract}). In contrast, we have worked up to this point in the fundamental space.
The exertion of expressing $D$-terms via only their corresponding state VEVs is rewarded by a natural
and straightforward implementation of $F$-flatness. However, even this ostensibly simple contraction is complicated in practice by the presence of distinct vector orientations for the real and imaginary VEVs.
Having encountered severe obstacles in the transition to a purely fundamental description, this section will instead entertain something of a hybrid approach.
Pragmatism will here overshadow the desire for generality, and an attempt will be made to extract from the previous technology some minimal working procedures to ensure simultaneous $D$- and $F$-flatness. New constructions will be added as needed.

There are three key quantities of interest in our search. These are the commutators between rotation generators, the scalar product between $D$-terms in the adjoint space, and the fundamental space VEV contraction.
First, we will define here two fully general complex vectors to be used throughout the discussion. No restriction on the orthonormality of the constituent elements is assumed.
\beq
\ket{\alpha} \equiv (\vec{a} - i\vec{b})/\sqrt{2}\, , \,\,\,
\ket{\gamma} \equiv (\vec{c} - i\vec{d})/\sqrt{2}
\label{agvects}
\eeq 

Turning first to the commutator,
\beq
[M_{\vec{a}\vec{b}},M_{\vec{c}\vec{d}}] =
(i)^2 \left\{
(\vec{a}\otimes\vec{b} - \vec{b}\otimes\vec{a})\cdot
(\vec{c}\otimes\vec{d} - \vec{d}\otimes\vec{c})
-(\vec{c}\otimes\vec{d} - \vec{d}\otimes\vec{c})\cdot
(\vec{a}\otimes\vec{b} - \vec{b}\otimes\vec{a})
\right\}\, ,
\label{vectcomm}
\eeq
using the form of Eq.\ (\ref{vectgen}).
This simplifies cleanly to
\beq
[M_{\vec{a}\vec{b}},M_{\vec{a}\vec{b}}] =
i\left\{
(\vec{a}\cdot\vec{d})M_{\vec{b}\vec{c}}
+(\vec{b}\cdot\vec{c})M_{\vec{a}\vec{d}}
-(\vec{a}\cdot\vec{c})M_{\vec{b}\vec{d}}
-(\vec{b}\cdot\vec{d})M_{\vec{a}\vec{c}}
\right\}\, ,
\label{vectrotalg}
\eeq
which is equivalent to Eq.\ (\ref{rotalg}) in the orthonormal limit.
This commutator vanishes if and only if the two rotations are fully disentangled, \ie the rotation planes have a null intersection such that ($\vec{a}$,$\vec{b}$) are mutually orthogonal to ($\vec{c}$,$\vec{d}$).
As before, the case that ($\vec{a}$,$\vec{b}$)\footnote{Or, equivalently, ($\vec{c}$,$\vec{d}$).} are themselves (anti)parallel is trivial, with $M_{\vec{a}\vec{b}} = 0$. 

Next, we will examine the adjoint space contraction between the $D$-terms in correspondence to each of $\ket{\alpha}$ and $\ket{\gamma}$, denoted henceforth as $(\alpha\cdot\gamma)_A$.
As established in Eq.\ (\ref{dtermgen}), the needed adjoint space `expectation values' are realized as the members of an antisymmetric matrix functionally identical, modulo the imaginary factor and a possible scale, to the rotation generator which would interpolate between vectors in the related fundamental space plane.
To take the scalar product in this adjoint space, we must sum over the product of corresponding pairs between the two $D$-terms, including each of the $m(m-1)/2$ unique basis members one time.
In terms of the provided form, this is realized as the trace of a matrix multiplication. 
\bea
(\alpha\cdot\gamma)_A &=&
-\frac{\alpha_{ij}\gamma_{ji}}{2} \nonumber\\
&=&
-1/2Tr\left\{
(\vec{a}\otimes\vec{b} - \vec{b}\otimes\vec{a})\cdot
(\vec{c}\otimes\vec{d} - \vec{d}\otimes\vec{c})
\right\}
\label{adjproduct} \\
&=&
-1/2Tr\left\{
(\vec{b}\cdot\vec{c})\vec{a}\otimes\vec{d} +
(\vec{a}\cdot\vec{d})\vec{b}\otimes\vec{c} -
(\vec{b}\cdot\vec{d})\vec{a}\otimes\vec{c} -
(\vec{a}\cdot\vec{c})\vec{b}\otimes\vec{d}
\right\}
\nonumber
\eea
A short diversion is in order here to investigate what is meant by the trace in this language. Recall that the notation $\vec{a}\otimes\vec{b}$ simply signifies the matrix constructed by the `outer' product of the vectors $\vec{a}$ and $\vec{b}$. Thus:
\bea
Tr(\vec{a}\otimes\vec{b}) &\equiv&
\left(
\begin{array}{cccc}
a_1 b_1 & a_1 b_2 & a_1 b_3 & \\
a_2 b_1 & a_2 b_2 & a_2 b_3 & \\
a_3 b_1 & a_3 b_2 & a_3 b_3 & \\
&&& \ddots
\end{array}
\right) \nonumber \\
&=& a_1 b_1 + a_2 b_2 + a_3 b_3 + \cdots \equiv
\vec{a}\cdot\vec{b}
\label{tracescp}
\eea
Armed with this knowledge, the expression from Eqs. (\ref{adjproduct}) reduces nicely.
\beq
(\alpha\cdot\gamma)_A =
(\vec{a}\cdot\vec{c})(\vec{b}\cdot\vec{d})
 - (\vec{b}\cdot\vec{c})(\vec{a}\cdot\vec{d})
\label{adjprodsm}
\eeq
The next question of relevance is whether any geometrically intuitive representation is possible.
We will make here a separation of the vectors $\vec{c}$ and $\vec{d}$ into two sections each, representing portions within and orthogonal to the $(\vec{a},\vec{b})$ plane.
\beq
\vec{c} \equiv \vec{c}_{\parallel} + \vec{c}_{\perp} \,\, , \,\,
\vec{d} \equiv \vec{d}_{\parallel} + \vec{d}_{\perp}
\label{othogcd}
\eeq
The orthogonal portions $(\vec{c}_{\perp},\vec{d}_{\perp})$ trivially factor out of Eq. (\ref{adjprodsm}), reducing the analysis to a single plane.
Furthermore, it is noted that the presence of all four vectors in both terms of this difference allow the extraction of an overall scale factor, leaving us to contend only with angles of orientation.
\bea
(\alpha\cdot\gamma)_A &\Rightarrow&
\vert a\,b\,c_{\parallel}d_{\parallel} \vert
\left\{
\cos(\delta)\cos(\delta+\gamma-\alpha) -
\cos(\alpha-\delta)\cos(\delta+\gamma)
\right\} \nonumber \\
&=& \vert a\,b\,c_{\parallel}d_{\parallel} \vert \sin(\alpha)\sin(\gamma)
\label{adjprodgeom}
\eea
The angle\footnote{When used in a trigonometric context a symbol such as $\alpha$ will designate the angle of separation between the real and imaginary constituents of the corresponding VEV state $\ket{\alpha}$.} $\alpha$ separates $(\vec{a},\vec{b})$, while $\gamma$ splits $(\vec{c}_{\parallel},\vec{d}_{\parallel})$, and $\delta$ is the angle between the vectors $(\vec{a},\vec{c}_{\parallel})$.
Basic trigonometric relations lead to the quite concise final result of Eq. (\ref{adjprodgeom}), which has also a pleasing interpretation. 
Each of the pairs $(\vec{a},\vec{b})$ and $(\vec{c},\vec{d})$ correspond to an area within their plane constructed by completing the parallelogram which contains the vector pair as edges.
Eq. (\ref{adjprodgeom}) is the product of these areas, including only the portion which they project onto each other.
As is required, for the case of $\ket{\alpha} = \ket{\gamma}$, this result reduces to the square of Eq. (\ref{Dmag})\footnote{Note for comparison that $R \equiv \frac{a}{\sqrt{2}}$ and $I \equiv -\frac{b}{\sqrt{2}}$}.

We note that there are then three distinct mechanisms accessible for tuning the value of an adjoint scalar contraction.
Firstly, the relative plane orientations can be tilted to effect a lesser or greater area of projection.
It is clear that if only {\it one} of $(\vec{c}_{\parallel},\vec{d}_{\parallel})$ vanishes, then the product is null.
The condition $(\alpha\cdot\gamma)_A = 0$ is thus {\it weaker} than the statement $[M_{\vec{a}\vec{b}},M_{\vec{c}\vec{d}}] = 0$, which requires complete independence of the two planes. 
Secondly, the internal angle between the real and imaginary portions of a single state can be adjusted to shorten or lengthen its overall $D$-term scale.
Thirdly, one may of course consider rescaling the magnitude of the coefficients $\vert a\,b\vert$, although the combination $(a^2 + b^2)/2 = |\ket{\alpha}|^2$ is generally constrained in units of the squared FI-scale.
In keeping with the discussion leading up to Eq. (\ref{dtermgen}), these last two scenarios have the benefit of leaving {\it intact} the adjoint space {\it orientation}.
The maximum adjoint space extension occurs when $\vert a \vert = \vert b \vert$ and $\vec{a}\cdot\vec{b}=0$, in which case $\vert D \vert = a^2 = |\ket{\alpha}|^2$, which is proportional to the corresponding integral multiple of the fundamental scale.

The final quantity of interest for additional study here is the $SO(2n)$ invariant contraction in the fundamental space.
This is simply the standard orthogonal inner product.
\bea
(\alpha \cdot \gamma)_F &\equiv& {\alpha}^T\gamma \nonumber \\
&\equiv& \frac{1}{2}(\vec{a}\cdot\vec{c} - \vec{b}\cdot\vec{d}) - \frac{i}{2}(\vec{b}\cdot\vec{c} + \vec{a}\cdot\vec{d})
\label{prodfund}
\eea

\section{Concluding Remarks}

Free Fermionic constructions require assignment of a non-trivial vacuum, oriented in
the field space to preserve a flat scalar potential, in order to trim field content,
provide masses and cancel the FI anomaly.
We have observed the emergence of new techniques for the removal of
dangerous terms from $\vev{W}$ and from $\vev{F}$ containing non-Abelian fields of the
Lie groups $SU(2)$ and $SO(2n)$, taking the stance that a geometrical interpretation
of the adjoint VEV representation will aid visual intuition of the model builder.
In Table 1B of \cite{Cleaver:2001ab},
four of our flat directions are lifted to all order by the vanishing of terms with
more than two \NA fields. As a second example, flatness of the phenomenologically preferred
directions of \cite{Perkins:2005zh} was lifted from $11^{\rm th}$ order to beyond $17^{\rm th}$ via 
an $SU(3)$ triplet/anti-triplet generalization of the $SU(2)$ self-cancellation example presented 
in section \ref{NASC} herein. 

One research path suggested by the partial success of these flat directions is investigation of non-stringently
flat directions for the FNY model that are flat to a finite order due to 
cancellation between various components in an $F$-term. 
While the simpler restriction of purely real state VEVs ensures perfect compliance with D-flatness,
a benefit of the more difficult case of non-trivial $D$-flatness
may be that using generalized vectors can force flatness to be preserved order by order.
So when flatness eventually fails at some high order, a natural scale might emerge at which supersymmetry is broken.
The non-trivial complex VEVs will likely be especially important when ratios between coupling constants
of differing components within an $F$-term are imaginary or complex.
It remains to be seen whether more attractive phenomenology with improved mass terms etc.\ will emerge.
Further, our discovery of a Minimal Standard Heterotic String Model in the 
neighborhood of the string/M-theory parameter space allowing  
free-fermionic description strongly suggests a search for 
further, perhaps more phenomenologically realistic, Minimal Standard Heterotic String Models in this
region.
It may also be valuable in time to extend the brief examples studied for application of
non-Abelian VEVs, particularly within the existing flipped $SU(5)$ Heterotic construction.
This we leave for future research.

The necessity of non-stringent $F$-flatness was also shown for free fermionic
flipped $SU(5)$ models \cite{Cleaver:2000sc} 
and for possible heterotic string realization of optical unification 
\cite{Cleaver:2002qc,Perkins:2003tb,Perkins:2005zh}.
Related classificiation of flipped $SU(5)$ finite-order directions, 
$F$-flat to at least $17^{\rm th}$ order (as required for viable 
observable-sector SUSY-beaking) is thus underway. An parallel investigation for
the optical unification model will then follow.
For the Flipped SU(5) model, special attention
is being given to directions providing the best three-generation and Higgs phenomenology \cite{cenp1}.
These latter directions have been shown to be flat up to {\it at least} 6th order and in some cases 
up to {\it at least} 10th order \cite{Antoniadis:1987dx}.
Additional VEV constraints necessitated by flatness up to at least $17^{\rm th}$ order are being determined
\cite{cenp1}. 

Geometrical interpretation of non-Abelian flat directions may also find relevant application through the recent conjecture
of equivalence between $D$-term strings and $D_{1+q}$-branes of Type II theory wrapped on a $q$-cycle \cite{Dvali:2003zh}. 
This conjectured connection should allow many unknown properties of brane-anti-brane systems to be interpreted in the 
language of 4D supergravity $D$-terms \cite{Binetruy:2004hh} (and vice-versa). In particular, the energy of an unstable 
$D_{3+q}- \bar{D}_{3+q}$ system can be viewed as FI $D$-term energy. The decay of this system to a stable $D_{1+q}$ 
string then corresponds to the appearance of a flat direction VEV to cancel the FI term. It has been argued that the 
geometry of the restrictions to the parameter space of flat direction VEVs of the $D$-term string relates to the 
shifting of a Ramond-Ramond axion field on the branes under the anomalous $U(1)$ symmetry \cite{Binetruy:2004hh}.
Thus, knowledge of the geometry of the flat direction ``landscape'' of a $D$-term string model could yield information 
about the dual brane model.

%% file: append.tex
\section{Appendix}
\noindent
Presented here is the choice of $SO(6)$ Matrices in the Original and Diagonal bases
\footnote{\lowercase{\footnotesize{\uppercase{T}he symbols $(+,-,\times)$ describe the raising, lowering or null role of the operators $\uppercase{E}_i$ in each of the three selected diagonal sectors.}}}.

\[
\begin{tiny}
\begin{array}{@{\hspace{-16pt}}c@{\hspace{8pt}}c@{\hspace{8pt}}c@{\hspace{0pt}}}
{\large M_{12}}
&{\large M_{34}}
&{\large M_{56}}
\\
\left(\begin{array}
{@{\hspace{3pt}}c@{\hspace{4pt}}c@{\hspace{4pt}}c@{\hspace{4pt}}c@{\hspace{4pt}}c@{\hspace{4pt}}c@{\hspace{3pt}}}
	 0 &  i &  0 &  0 &  0 &  0\\
	-i &  0 &  0 &  0 &  0 &  0\\
	 0 &  0 &  0 &  0 &  0 &  0\\
	 0 &  0 &  0 &  0 &  0 &  0\\
	 0 &  0 &  0 &  0 &  0 &  0\\
	 0 &  0 &  0 &  0 &  0 &  0\end{array}\right),
\left(\begin{array}
{@{\hspace{3pt}}c@{\hspace{4pt}}c@{\hspace{4pt}}c@{\hspace{4pt}}c@{\hspace{4pt}}c@{\hspace{4pt}}c@{\hspace{3pt}}}
	 1 &  0 &  0 &  0 &  0 &  0\\
	 0 & -1 &  0 &  0 &  0 &  0\\
	 0 &  0 &  0 &  0 &  0 &  0\\
	 0 &  0 &  0 &  0 &  0 &  0\\
	 0 &  0 &  0 &  0 &  0 &  0\\
	 0 &  0 &  0 &  0 &  0 &  0\end{array}\right);
&\left(\begin{array}
{@{\hspace{3pt}}c@{\hspace{4pt}}c@{\hspace{4pt}}c@{\hspace{4pt}}c@{\hspace{4pt}}c@{\hspace{4pt}}c@{\hspace{3pt}}}
	 0 &  0 &  0 &  0 &  0 &  0\\
	 0 &  0 &  0 &  0 &  0 &  0\\
	 0 &  0 &  0 &  i &  0 &  0\\
	 0 &  0 & -i &  0 &  0 &  0\\
	 0 &  0 &  0 &  0 &  0 &  0\\
	 0 &  0 &  0 &  0 &  0 &  0\end{array}\right),
\left(\begin{array}
{@{\hspace{3pt}}c@{\hspace{4pt}}c@{\hspace{4pt}}c@{\hspace{4pt}}c@{\hspace{4pt}}c@{\hspace{4pt}}c@{\hspace{3pt}}}
	 0 &  0 &  0 &  0 &  0 &  0\\
	 0 &  0 &  0 &  0 &  0 &  0\\
	 0 &  0 &  1 &  0 &  0 &  0\\
	 0 &  0 &  0 & -1 &  0 &  0\\
	 0 &  0 &  0 &  0 &  0 &  0\\
	 0 &  0 &  0 &  0 &  0 &  0\end{array}\right);
&\left(\begin{array}
{@{\hspace{3pt}}c@{\hspace{4pt}}c@{\hspace{4pt}}c@{\hspace{4pt}}c@{\hspace{4pt}}c@{\hspace{4pt}}c@{\hspace{3pt}}}
	 0 &  0 &  0 &  0 &  0 &  0\\
	 0 &  0 &  0 &  0 &  0 &  0\\
	 0 &  0 &  0 &  0 &  0 &  0\\
	 0 &  0 &  0 &  0 &  0 &  0\\
	 0 &  0 &  0 &  0 &  0 &  i\\
	 0 &  0 &  0 &  0 & -i &  0\end{array}\right),
\left(\begin{array}
{@{\hspace{3pt}}c@{\hspace{4pt}}c@{\hspace{4pt}}c@{\hspace{4pt}}c@{\hspace{4pt}}c@{\hspace{4pt}}c@{\hspace{3pt}}}
	 0 &  0 &  0 &  0 &  0 &  0\\
	 0 &  0 &  0 &  0 &  0 &  0\\
	 0 &  0 &  0 &  0 &  0 &  0\\
	 0 &  0 &  0 &  0 &  0 &  0\\
	 0 &  0 &  0 &  0 &  1 &  0\\
	 0 &  0 &  0 &  0 &  0 & -1\end{array}\right);
\end{array}
\end{tiny}
\]

\[
\begin{tiny}
\begin{array}
{@{\hspace{-4pt}}c@{\hspace{2pt}}c@{\hspace{3pt}}c@{\hspace{3pt}}c@{\hspace{2pt}}c
@{\hspace{8pt}}c@{\hspace{2pt}}c@{\hspace{3pt}}c@{\hspace{3pt}}c@{\hspace{2pt}}c@{\hspace{0pt}}}
{\large E_{1}}
&\left[\begin{array}{@{\hspace{2pt}}c@{\hspace{2pt}}}+\\+\\\times\end{array}\right]
&\equiv
&\frac{1}{2}\left(\begin{array}
{@{\hspace{3pt}}c@{\hspace{4pt}}c@{\hspace{4pt}}c@{\hspace{4pt}}c@{\hspace{4pt}}c@{\hspace{4pt}}c@{\hspace{3pt}}}
	 0 &  0 & -1 &  i &  0 &  0\\
	 0 &  0 &  i &  1 &  0 &  0\\
	 1 & -i &  0 &  0 &  0 &  0\\
	-i & -1 &  0 &  0 &  0 &  0\\
	 0 &  0 &  0 &  0 &  0 &  0\\
	 0 &  0 &  0 &  0 &  0 &  0\end{array}\right),
&\left(\begin{array}
{@{\hspace{3pt}}c@{\hspace{4pt}}c@{\hspace{4pt}}c@{\hspace{4pt}}c@{\hspace{4pt}}c@{\hspace{4pt}}c@{\hspace{3pt}}}
	 0 &  0 &  0 & -1 &  0 &  0\\
	 0 &  0 &  0 &  0 &  0 &  0\\
	 0 &  1 &  0 &  0 &  0 &  0\\
	 0 &  0 &  0 &  0 &  0 &  0\\
	 0 &  0 &  0 &  0 &  0 &  0\\
	 0 &  0 &  0 &  0 &  0 &  0\end{array}\right);
&{\large E_{2}}
&\left[\begin{array}{@{\hspace{2pt}}c@{\hspace{2pt}}}-\\-\\\times\end{array}\right]
&\equiv
&\frac{1}{2}\left(\begin{array}
{@{\hspace{3pt}}c@{\hspace{4pt}}c@{\hspace{4pt}}c@{\hspace{4pt}}c@{\hspace{4pt}}c@{\hspace{4pt}}c@{\hspace{3pt}}}
	 0 &  0 &  1 &  i &  0 &  0\\
	 0 &  0 &  i & -1 &  0 &  0\\
	-1 & -i &  0 &  0 &  0 &  0\\
	-i &  1 &  0 &  0 &  0 &  0\\
	 0 &  0 &  0 &  0 &  0 &  0\\
	 0 &  0 &  0 &  0 &  0 &  0\end{array}\right),
&\left(\begin{array}
{@{\hspace{3pt}}c@{\hspace{4pt}}c@{\hspace{4pt}}c@{\hspace{4pt}}c@{\hspace{4pt}}c@{\hspace{4pt}}c@{\hspace{3pt}}}
	 0 &  0 &  0 &  0 &  0 &  0\\
	 0 &  0 &  1 &  0 &  0 &  0\\
	 0 &  0 &  0 &  0 &  0 &  0\\
	-1 &  0 &  0 &  0 &  0 &  0\\
	 0 &  0 &  0 &  0 &  0 &  0\\
	 0 &  0 &  0 &  0 &  0 &  0\end{array}\right);
\\
\\
{\large E_{3}}
&\left[\begin{array}{@{\hspace{2pt}}c@{\hspace{2pt}}}+\\-\\\times\end{array}\right]
&\equiv
&\frac{1}{2}\left(\begin{array}
{@{\hspace{3pt}}c@{\hspace{4pt}}c@{\hspace{4pt}}c@{\hspace{4pt}}c@{\hspace{4pt}}c@{\hspace{4pt}}c@{\hspace{3pt}}}
	 0 &  0 &  1 &  i &  0 &  0\\
	 0 &  0 & -i &  1 &  0 &  0\\
	-1 &  i &  0 &  0 &  0 &  0\\
	-i & -1 &  0 &  0 &  0 &  0\\
	 0 &  0 &  0 &  0 &  0 &  0\\
	 0 &  0 &  0 &  0 &  0 &  0\end{array}\right),
&\left(\begin{array}
{@{\hspace{3pt}}c@{\hspace{4pt}}c@{\hspace{4pt}}c@{\hspace{4pt}}c@{\hspace{4pt}}c@{\hspace{4pt}}c@{\hspace{3pt}}}
	 0 &  0 &  1 &  0 &  0 &  0\\
	 0 &  0 &  0 &  0 &  0 &  0\\
	 0 &  0 &  0 &  0 &  0 &  0\\
	 0 & -1 &  0 &  0 &  0 &  0\\
	 0 &  0 &  0 &  0 &  0 &  0\\
	 0 &  0 &  0 &  0 &  0 &  0\end{array}\right);
&{\large E_{4}}
&\left[\begin{array}{@{\hspace{2pt}}c@{\hspace{2pt}}}-\\+\\\times\end{array}\right]
&\equiv
&\frac{1}{2}\left(\begin{array}
{@{\hspace{3pt}}c@{\hspace{4pt}}c@{\hspace{4pt}}c@{\hspace{4pt}}c@{\hspace{4pt}}c@{\hspace{4pt}}c@{\hspace{3pt}}}
	 0 &  0 & -1 &  i &  0 &  0\\
	 0 &  0 & -i & -1 &  0 &  0\\
	 1 &  i &  0 &  0 &  0 &  0\\
	-i &  1 &  0 &  0 &  0 &  0\\
	 0 &  0 &  0 &  0 &  0 &  0\\
	 0 &  0 &  0 &  0 &  0 &  0\end{array}\right),
&\left(\begin{array}
{@{\hspace{3pt}}c@{\hspace{4pt}}c@{\hspace{4pt}}c@{\hspace{4pt}}c@{\hspace{4pt}}c@{\hspace{4pt}}c@{\hspace{3pt}}}
	 0 &  0 &  0 &  0 &  0 &  0\\
	 0 &  0 &  0 & -1 &  0 &  0\\
	 1 &  0 &  0 &  0 &  0 &  0\\
	 0 &  0 &  0 &  0 &  0 &  0\\
	 0 &  0 &  0 &  0 &  0 &  0\\
	 0 &  0 &  0 &  0 &  0 &  0\end{array}\right);
\\
\\
{\large E_{5}}
&\left[\begin{array}{@{\hspace{2pt}}c@{\hspace{2pt}}}+\\\times\\+\end{array}\right]
&\equiv
&\frac{1}{2}\left(\begin{array}
{@{\hspace{3pt}}c@{\hspace{4pt}}c@{\hspace{4pt}}c@{\hspace{4pt}}c@{\hspace{4pt}}c@{\hspace{4pt}}c@{\hspace{3pt}}}
	 0 &  0 &  0 &  0 & -i & -1\\
	 0 &  0 &  0 &  0 & -1 &  i\\
	 0 &  0 &  0 &  0 &  0 &  0\\
	 0 &  0 &  0 &  0 &  0 &  0\\
	 i &  1 &  0 &  0 &  0 &  0\\
	 1 & -i &  0 &  0 &  0 &  0\end{array}\right),
&\left(\begin{array}
{@{\hspace{3pt}}c@{\hspace{4pt}}c@{\hspace{4pt}}c@{\hspace{4pt}}c@{\hspace{4pt}}c@{\hspace{4pt}}c@{\hspace{3pt}}}
	 0 &  0 &  0 &  0 &  0 & -i\\
	 0 &  0 &  0 &  0 &  0 &  0\\
	 0 &  0 &  0 &  0 &  0 &  0\\
	 0 &  0 &  0 &  0 &  0 &  0\\
	 0 &  i &  0 &  0 &  0 &  0\\
	 0 &  0 &  0 &  0 &  0 &  0\end{array}\right);
&{\large E_{6}}
&\left[\begin{array}{@{\hspace{2pt}}c@{\hspace{2pt}}}-\\\times\\-\end{array}\right]
&\equiv
&\frac{1}{2}\left(\begin{array}
{@{\hspace{3pt}}c@{\hspace{4pt}}c@{\hspace{4pt}}c@{\hspace{4pt}}c@{\hspace{4pt}}c@{\hspace{4pt}}c@{\hspace{3pt}}}
	 0 &  0 &  0 &  0 & -i &  1\\
	 0 &  0 &  0 &  0 &  1 &  i\\
	 0 &  0 &  0 &  0 &  0 &  0\\
	 0 &  0 &  0 &  0 &  0 &  0\\
	 i & -1 &  0 &  0 &  0 &  0\\
	-1 & -i &  0 &  0 &  0 &  0\end{array}\right),
&\left(\begin{array}
{@{\hspace{3pt}}c@{\hspace{4pt}}c@{\hspace{4pt}}c@{\hspace{4pt}}c@{\hspace{4pt}}c@{\hspace{4pt}}c@{\hspace{3pt}}}
	 0 &  0 &  0 &  0 &  0 &  0\\
	 0 &  0 &  0 &  0 & -i &  0\\
	 0 &  0 &  0 &  0 &  0 &  0\\
	 0 &  0 &  0 &  0 &  0 &  0\\
	 0 &  0 &  0 &  0 &  0 &  0\\
	 i &  0 &  0 &  0 &  0 &  0\end{array}\right);
\\
\\
{\large E_{7}}
&\left[\begin{array}{@{\hspace{2pt}}c@{\hspace{2pt}}}+\\\times\\-\end{array}\right]
&\equiv
&\frac{1}{2}\left(\begin{array}
{@{\hspace{3pt}}c@{\hspace{4pt}}c@{\hspace{4pt}}c@{\hspace{4pt}}c@{\hspace{4pt}}c@{\hspace{4pt}}c@{\hspace{3pt}}}
	 0 &  0 &  0 &  0 &  i & -1\\
	 0 &  0 &  0 &  0 &  1 &  i\\
	 0 &  0 &  0 &  0 &  0 &  0\\
	 0 &  0 &  0 &  0 &  0 &  0\\
	-i & -1 &  0 &  0 &  0 &  0\\
	 1 & -i &  0 &  0 &  0 &  0\end{array}\right),
&\left(\begin{array}
{@{\hspace{3pt}}c@{\hspace{4pt}}c@{\hspace{4pt}}c@{\hspace{4pt}}c@{\hspace{4pt}}c@{\hspace{4pt}}c@{\hspace{3pt}}}
	 0 &  0 &  0 &  0 &  i &  0\\
	 0 &  0 &  0 &  0 &  0 &  0\\
	 0 &  0 &  0 &  0 &  0 &  0\\
	 0 &  0 &  0 &  0 &  0 &  0\\
	 0 &  0 &  0 &  0 &  0 &  0\\
	 0 & -i &  0 &  0 &  0 &  0\end{array}\right);
&{\large E_{8}}
&\left[\begin{array}{@{\hspace{2pt}}c@{\hspace{2pt}}}-\\\times\\+\end{array}\right]
&\equiv
&\frac{1}{2}\left(\begin{array}
{@{\hspace{3pt}}c@{\hspace{4pt}}c@{\hspace{4pt}}c@{\hspace{4pt}}c@{\hspace{4pt}}c@{\hspace{4pt}}c@{\hspace{3pt}}}
	 0 &  0 &  0 &  0 &  i &  1\\
	 0 &  0 &  0 &  0 & -1 &  i\\
	 0 &  0 &  0 &  0 &  0 &  0\\
	 0 &  0 &  0 &  0 &  0 &  0\\
	-i &  1 &  0 &  0 &  0 &  0\\
	-1 & -i &  0 &  0 &  0 &  0\end{array}\right),
&\left(\begin{array}
{@{\hspace{3pt}}c@{\hspace{4pt}}c@{\hspace{4pt}}c@{\hspace{4pt}}c@{\hspace{4pt}}c@{\hspace{4pt}}c@{\hspace{3pt}}}
	 0 &  0 &  0 &  0 &  0 &  0\\
	 0 &  0 &  0 &  0 &  0 &  i\\
	 0 &  0 &  0 &  0 &  0 &  0\\
	 0 &  0 &  0 &  0 &  0 &  0\\
	-i &  0 &  0 &  0 &  0 &  0\\
	 0 &  0 &  0 &  0 &  0 &  0\end{array}\right);
\\
\\
{\large E_{9}}
&\left[\begin{array}{@{\hspace{2pt}}c@{\hspace{2pt}}}\times\\+\\+\end{array}\right]
&\equiv
&\frac{1}{2}\left(\begin{array}
{@{\hspace{3pt}}c@{\hspace{4pt}}c@{\hspace{4pt}}c@{\hspace{4pt}}c@{\hspace{4pt}}c@{\hspace{4pt}}c@{\hspace{3pt}}}
	 0 &  0 &  0 &  0 &  0 &  0\\
	 0 &  0 &  0 &  0 &  0 &  0\\
	 0 &  0 &  0 &  0 & -1 &  i\\
	 0 &  0 &  0 &  0 &  i &  1\\
	 0 &  0 &  1 & -i &  0 &  0\\
	 0 &  0 & -i & -1 &  0 &  0\end{array}\right),
&\left(\begin{array}
{@{\hspace{3pt}}c@{\hspace{4pt}}c@{\hspace{4pt}}c@{\hspace{4pt}}c@{\hspace{4pt}}c@{\hspace{4pt}}c@{\hspace{3pt}}}
	 0 &  0 &  0 &  0 &  0 &  0\\
	 0 &  0 &  0 &  0 &  0 &  0\\
	 0 &  0 &  0 &  0 &  0 & -1\\
	 0 &  0 &  0 &  0 &  0 &  0\\
	 0 &  0 &  0 &  1 &  0 &  0\\
	 0 &  0 &  0 &  0 &  0 &  0\end{array}\right);
&{\large E_{10}}
&\left[\begin{array}{@{\hspace{2pt}}c@{\hspace{2pt}}}\times\\-\\-\end{array}\right]
&\equiv
&\frac{1}{2}\left(\begin{array}
{@{\hspace{3pt}}c@{\hspace{4pt}}c@{\hspace{4pt}}c@{\hspace{4pt}}c@{\hspace{4pt}}c@{\hspace{4pt}}c@{\hspace{3pt}}}
	 0 &  0 &  0 &  0 &  0 &  0\\
	 0 &  0 &  0 &  0 &  0 &  0\\
	 0 &  0 &  0 &  0 &  1 &  i\\
	 0 &  0 &  0 &  0 &  i & -1\\
	 0 &  0 & -1 & -i &  0 &  0\\
	 0 &  0 & -i &  1 &  0 &  0\end{array}\right),
&\left(\begin{array}
{@{\hspace{3pt}}c@{\hspace{4pt}}c@{\hspace{4pt}}c@{\hspace{4pt}}c@{\hspace{4pt}}c@{\hspace{4pt}}c@{\hspace{3pt}}}
	 0 &  0 &  0 &  0 &  0 &  0\\
	 0 &  0 &  0 &  0 &  0 &  0\\
	 0 &  0 &  0 &  0 &  0 &  0\\
	 0 &  0 &  0 &  0 &  1 &  0\\
	 0 &  0 &  0 &  0 &  0 &  0\\
	 0 &  0 & -1 &  0 &  0 &  0\end{array}\right);
\\
\\
{\large E_{11}}
&\left[\begin{array}{@{\hspace{2pt}}c@{\hspace{2pt}}}\times\\+\\-\end{array}\right]
&\equiv
&\frac{1}{2}\left(\begin{array}
{@{\hspace{3pt}}c@{\hspace{4pt}}c@{\hspace{4pt}}c@{\hspace{4pt}}c@{\hspace{4pt}}c@{\hspace{4pt}}c@{\hspace{3pt}}}
	 0 &  0 &  0 &  0 &  0 &  0\\
	 0 &  0 &  0 &  0 &  0 &  0\\
	 0 &  0 &  0 &  0 & -1 & -i\\
	 0 &  0 &  0 &  0 &  i & -1\\
	 0 &  0 &  1 & -i &  0 &  0\\
	 0 &  0 &  i &  1 &  0 &  0\end{array}\right),
&\left(\begin{array}
{@{\hspace{3pt}}c@{\hspace{4pt}}c@{\hspace{4pt}}c@{\hspace{4pt}}c@{\hspace{4pt}}c@{\hspace{4pt}}c@{\hspace{3pt}}}
	 0 &  0 &  0 &  0 &  0 &  0\\
	 0 &  0 &  0 &  0 &  0 &  0\\
	 0 &  0 &  0 &  0 & -1 &  0\\
	 0 &  0 &  0 &  0 &  0 &  0\\
	 0 &  0 &  0 &  0 &  0 &  0\\
	 0 &  0 &  0 &  1 &  0 &  0\end{array}\right);
&{\large E_{12}}
&\left[\begin{array}{@{\hspace{2pt}}c@{\hspace{2pt}}}\times\\-\\+\end{array}\right]
&\equiv
&\frac{1}{2}\left(\begin{array}
{@{\hspace{3pt}}c@{\hspace{4pt}}c@{\hspace{4pt}}c@{\hspace{4pt}}c@{\hspace{4pt}}c@{\hspace{4pt}}c@{\hspace{3pt}}}
	 0 &  0 &  0 &  0 &  0 &  0\\
	 0 &  0 &  0 &  0 &  0 &  0\\
	 0 &  0 &  0 &  0 &  1 & -i\\
	 0 &  0 &  0 &  0 &  i &  1\\
	 0 &  0 & -1 & -i &  0 &  0\\
	 0 &  0 &  i & -1 &  0 &  0\end{array}\right),
&\left(\begin{array}
{@{\hspace{3pt}}c@{\hspace{4pt}}c@{\hspace{4pt}}c@{\hspace{4pt}}c@{\hspace{4pt}}c@{\hspace{4pt}}c@{\hspace{3pt}}}
	 0 &  0 &  0 &  0 &  0 &  0\\
	 0 &  0 &  0 &  0 &  0 &  0\\
	 0 &  0 &  0 &  0 &  0 &  0\\
	 0 &  0 &  0 &  0 &  0 &  1\\
	 0 &  0 & -1 &  0 &  0 &  0\\
	 0 &  0 &  0 &  0 &  0 &  0\end{array}\right);
\end{array}
\end{tiny}
\]